\documentclass[aps,prb,twocolumn]{revtex4-2}

\usepackage{graphicx}
\usepackage{upgreek}
\usepackage{amsmath}
\usepackage{amssymb}
\usepackage{dcolumn}
\usepackage{chngpage}

\usepackage{CJK}

\usepackage{graphicx}
\usepackage{dcolumn}
\usepackage{bm}
\usepackage{amssymb} 
\usepackage{subfigure}
\usepackage{color}
\usepackage{epstopdf}
\usepackage{tabularx}

\begin{document}

\title{$^{139}$La NMR investigation of the interplay between the lattice, charge, and spin dynamics in charge ordered high $T_c$ cuprate La$_{1.875}$Ba$_{0.125}$CuO$_{4}$}

\author{P. M. Singer$^{1}$}
\author{A. Arsenault$^{2}$}
\author{T. Imai$^{2}$}
\author{M. Fujita$^{3}$}
\affiliation{$^{1}$Department of Chemical and Biomolecular Engineering, Rice University, 6100 Main St., Houston, TX 77005-1892, United States}
\affiliation{$^{2}$Department of Physics and Astronomy, McMaster University, Hamilton, Ontario, L8S 4M1, Canada}
\affiliation{$^{3}$Institute for Materials Research, Tohoku University, Sendai 980-8577, Japan}





\date{\today}

\begin{abstract}
We investigate the interplay between the lattice, charge, and spin dynamics in charge ordered high $T_c$ cuprate La$_{1.875}$Ba$_{0.125}$CuO$_{4}$ ($T_{c} =4$~K) based on the inverse Laplace transform (ILT) analysis of the $^{139}$La nuclear spin-lattice relaxation rate $1/T_1$ (dubbed ILTT$_{1}$ analysis here after). A major thrust of the ILTT$_{1}$ analysis is that one can deduce the probability density function $P(1/T_1)$ of distributed $1/T_1$. We demonstrate that $1/T_{1}^{lm}$, defined as the log-mean (i.e. the center of gravity on a logarithmic scale) of $P(1/T_1)$, can be well approximated by $1/T_{1}^{str}$ deduced from the phenomenological stretched fit, however, $P(1/T_1)$ can provide much richer insight into how the lattice, charge, and spin fluctuations and their distribution develop near and below the long range charge order at $T_{charge} \sim 54$~K. Upon entering the charge ordered state, a divergent increase of $1/T_{1}^{lm}$ toward the spin ordering at $T_{spin}^{\mu SR} \simeq 35$~K is accompanied by an asymmetric broadening of $P(1/T_1)$.   Even deep inside the charge ordered state, $1/T_{1}$ at a gradually diminishing fraction of $^{139}$La sites continues to slow down as temperature is lowered, as expected for canonical superconducting CuO$_2$ planes without enhanced spin fluctuations.  The fraction of such canonical $^{139}$La sites almost disappears by $\simeq 40$~K.  In contrast, nearly a half of the $^{139}$La sites in La$_{1.885}$Sr$_{0.115}$CuO$_{4}$ ($T_{charge} \simeq 80$~K) still exhibits the canonical behavior without enhanced spin fluctuations even near its $T_{c} = 31$~K.  These contrasting behaviors explain why superconductivity in La$_{1.875}$Ba$_{0.125}$CuO$_{4}$ is more strongly suppressed than in La$_{1.885}$Sr$_{0.115}$CuO$_{4}$ despite the lower onset temperature of the charge order. 
\end{abstract}


\maketitle


\section{General Introduction}
In La$_{2-x}$Sr$_{x}$CuO$_{4}$, the high temperature superconducting phase with the critical temperature as high as $T_{c} =38$~K manifests itself after the high temperature tetragonal (HTT) to low temperature orthorhombic (LTO) structural phase transition takes place around $\simeq 200$~K. In contrast, La$_{2-x}$Ba$_{x}$CuO$_{4}$ undergoes an additional structural phase transition from the LTO to the low temperature tetragonal (LTT) phase below $T_{LTO-LTT}\simeq 60$~K, and $T_c$ in the LTT structure is anomalously suppressed to as low as $T_{c}\simeq 4$~K near the magic composition with $x \simeq 1/8$ \cite{Axe1989}. $\mu$SR measurements uncovered the presence of spin order below $T_{spin}^{\mu SR}\simeq 35$~K in La$_{1.875}$Ba$_{0.125}$CuO$_{4}$ \cite{Luke1991}.  

Nd co-doping into La$_{2-x}$Sr$_{x}$CuO$_{4}$ also induces the same sequence of structural transitions from HTT to LTO, and then to LTT.  It was in La$_{1.48}$Nd$_{0.4}$Sr$_{0.12}$CuO$_{4}$ where Tranquada et al. \cite{Tranquada1995} discovered the charge order transition at $T_{charge} \simeq 60$~K in the LTT phase based on neutron diffraction measurements.  Subsequent NMR \cite{HuntPRL1999, HuntPRB2001} and neutron scattering experiments \cite{Fujita2004} showed that a charge order transition accompanies the LTO to LTT structural transition also in La$_{1.875}$Ba$_{0.125}$CuO$_{4}$. More recently, x-ray scattering experiments \cite{Thampy2017, MiaoPRX2019} revealed the presence of dynamic short range charge order in La$_{1.875}$Ba$_{0.125}$CuO$_{4}$ prior to the onset of long range charge order at $T_{charge} \simeq 54$~K \cite{MiaoPRX2019}.  

There has been a long history in the NMR investigation of the complicated behavior of La$_{2-x}$Ba$_{x}$CuO$_{4}$ \cite{ImaiJPSJ1990, TouLBCOT2, TouLBCOT1, Kumagai1994, GotoJPSJ1994, HuntPRL1999, HuntPRB2001, BaekLaT1PRB2015, Pelc2017}, yet our understanding of the interplay between the lattice, charge, and spin degrees of freedom near charge order is still far from complete. This is in part because the NMR spin-lattice relaxation rate $1/T_{1}$ develops a large distribution near and below $T_{charge}$, and the NMR community in condensed matter physics did not have the machinery to accurately probe the nature and extent of the distribution.  Furthermore, the phenomenological approach to deduce $1/T_1$ based on the stretched exponential fit of the nuclear spin recovery curve $M(t)$ cannot distinguish the fluctuations of the electric field gradient (EFG) and spins.

In this paper, we shed new light on the complex behavior of La$_{1.875}$Ba$_{0.125}$CuO$_{4}$ by analyzing the $1/T_{1}$ recovery curve $M(t)$ observed at the $^{139}$La sites based on the inverse Laplace transform (ILT) analysis techniques (dubbed {\it ILTT$_{1}$ analysis} hereafter). The ILT in the context of NMR has been conceptually known for some time \cite{JohnstonPRL2005}, and applied successfully in petrophysics \cite{venkataramanan:ieee2002,song:jmr2002,mitchell:PNMRS2012,singer:jcp2018,singer:jcp2018b} and condensed matter physics \cite{ArsenaultPRB2019,TakahashiPRX2019} by numerically inverting $M(t)$. A major thrust of the ILTT$_{1}$ analysis is that {\it one can generate the histogram of distributed $1/T_1$ without presuming any particular functional form of the density function $P(1/T_{1})$}, in addition to the log-mean $1/T_{1}^{lm}$ (i.e. the center of gravity on a logarithmic scale) of the distributed relaxation rate. We will demonstrate that slow lattice and/or charge fluctuations develop at the NMR frequency scales below $\simeq 80$~K, where the dynamic short range charge order develops \cite{MiaoPRX2019}. Comparison of $P(1/T_1)$ between La$_{1.875}$Ba$_{0.125}$CuO$_{4}$ ($T_{c} = 4$~K) and La$_{1.885}$Sr$_{0.115}$CuO$_{4}$ ($T_{c}\simeq 31$~K) also reveals a qualitative difference between the two. The volume fraction of the canonically superconducting domains in the CuO$_2$ planes without enhanced spin fluctuations is reduced to almost null in  La$_{1.875}$Ba$_{0.125}$CuO$_{4}$ below $\simeq 40$~K, whereas nearly a half of the volume still behaves as a canonical superconductor in La$_{1.885}$Sr$_{0.115}$CuO$_{4}$.

The rest of this paper is organized as follows.  In section II, we explain the key aspects of the ILT techniques and what the ILTT$_{1}$ analysis can (not) do. Section III outlines the experimental methods, and section IV discusses the results, followed by summary and conclusions in section V. 

\section{Inverse Laplace Transform of $M(t)$}
To measure the spin-lattice relaxation time $T_{1}$, one applies an inversion pulse and monitors the recovery curve $M(t)$ of the nuclear magnetization as a function of time $t$. In the simplest case of nuclear spin $I=1/2$ with a single, non-distributed value of $1/T_{1}$, the recovery obeys an exponential form as such:
\begin{equation}
M(t) = M_{0}-A\, e^{-t/T_{1}},	\label{eq:exp}
\end{equation} 
where $M_{0}$ is the saturated value of the nuclear magnetization, and $A$ ($\leq 2M_0$) represents the degree of inversion.  

Magnetic inhomogeneity of the sample results in a distribution of $1/T_1$, and $M(t)$ may exhibit a stretched exponential form under certain circumstances \cite{Itoh1986, Thayamballi1980, JohnstonPRL2005},
\begin{equation}
M(t) = M_{0}-A\,e^{-(t/T_{1}^{str})^{\beta}} , \label{eq:stretch}	
\end{equation} 
where $\beta$ is the stretched exponent less than 1. If $1/T_{1}$ has no distribution, $1/T_{1}^{str} = 1/T_{1}$ and $\beta =1$. For example, in the case of $1/T_1$ measured by $^{63}$Cu nuclear quadrupole resonance (NQR) techniques in a highly disordered YBa$_2$Cu$_3$O$_{6.9}$ sample ($T_{c}=92$~K), Eq. (\ref{eq:stretch}) worked well far below $T_c$ with the exponent $\beta = 1/2$\cite{ImaiJPSJ1988}, which is expected for diffusion-less, enhanced relaxation caused by defect spins \cite{Thayamballi1980, Itoh1986}. One can justify such a stretched fit, if $\ln[M_{0}-M(t)]$ vs. $t^{\beta}$ or $\ln[M_{0}-M(t)]$ vs. $\ln(t)$ yields a straight line \cite{Itoh1986, ImaiJPSJ1988}.

One needs to bear in mind that the distributed relaxation mechanisms do not always lead to the stretched form in Eq. (\ref{eq:stretch}). In fact, Eq. (\ref{eq:stretch}) failed for YBa$_2$Cu$_3$O$_{6.95}$ ($T_{c}=92$~K) with less magnetic defects \cite{ImaiJPSJ1988-2, ImaiPhysicaC1989}. Instead, a two component fit worked well, with two distinct values of fast and slow relaxation rates, $1/T_{1}^{fast}$ and $1/T_{1}^{slow}$:   
\begin{equation}
M(t) = M_{0}-A_{fast} e^{-t/T_{1}^{fast}} - A_{slow} e^{-t/T_{1}^{slow}}. \label{eq:FastSlow}
\end{equation}   
In this case, dilute defect spins affect only the nuclear spins in their vicinity, and the rest of the superconductor exhibit much longer, intrinsic $1/T_{1}^{slow}$. An analogous situation arises for $1/T_{1}$ measured for type II superconductors under the presence of vortex cores induced by an external magnetic field. Accordingly, in NMR research on superconductivity, it is usually $1/T_{1}^{slow}$ that researchers present far below $T_c$.

A dilemma arises if there is no clearcut justification for Eq. (\ref{eq:stretch}) or (\ref{eq:FastSlow}). In fact, in the case of $^{19}$F ($I=1/2$) NMR investigations of the diluted antiferromagnet Mn$_{1-x}$Zn$_{x}$F$_{2}$, the $\ln[M_{0}-M(t)]$ vs. $\ln(t)$ plot revealed that the experimental reality is the combination of Eq. (\ref{eq:stretch}) and Eq.(\ref{eq:FastSlow}) \cite{Itoh1986}. Furthermore, additional complications arise if $M(t)$ under consideration is not for $I=1/2$. For example, $^{139}$La is a spin $I=7/2$ nucleus. In an ideal case of non-distributed $1/T_1$, $M(t)$ for the $I_{z}=-1/2 \leftrightarrow +1/2$ (i.e. central) transition of $I=7/2$ can be written analytically as a linear combination of 4 normal modes \cite{Andrew1961, Narath1967}:
\begin{equation}
M(t) = M_0 - A \,\sum_{k=1}^4 p_k e^{-q_k t/T_{1}}, \label{eq:expLa}
\end{equation}
where $p_k = \{1/84,3/44,75/364,1225/1716\}$ and $q_k = \{1,6,15,28\}$, and $\sum_{k=1}^4 p_k = 1$ is normalized. In this case, the aforementioned log plot $\ln [M_{0}-M(t)]$ does not yield a straight line, even if there is no distribution in $1/T_1$. Then how should we choose and justify the appropriate fitting function for $M(t)$ when $1/T_1$ is distributed and the fit with Eq. (\ref{eq:expLa}) is unsatisfactory? Should we just stretch each exponential term as in Eq. (\ref{eq:stretch}), or consider two or more values of $1/T_1$ as in Eq. (\ref{eq:FastSlow})? Or a combination of both?

The ILTT$_{1}$ analysis technique tells us the answers to these questions. Regardless of the origin of the distribution of $1/T_1$ and its extent, one can actually {\it deduce} the probability density function $P(1/T_{1})$ of $1/T_{1}$ based on the inverse Laplace transform (ILT) of $M(t)$ \cite{SingerJCP2018}. The ILT consists of fitting $M(t)$ to a sum of exponentials with decay rate $1/T_{1j}$ and weight $P(1/T_{1j}) \geq 0$. For the simplest case of $I=1/2$, the discrete form of the ILT inverts for $P(1/T_{1})$ assuming the following expression for the experimental data $M(t)$ \cite{venkataramanan:ieee2002,song:jmr2002,mitchell:PNMRS2012,singer:jcp2018,singer:jcp2018b}:
\begin{align}
M(t) &= \sum_{j=1}^m\left[1-2 \,e^{-t/T_{1j}}\right]P\!\left(1/T_{1j}\right). \label{eq:ILTexp}
\end{align}  
For simplicity, we assumed perfect inversion (i.e. $A = 2M_{o}$).  The summation $\sum_{j=1}^m P(1/T_{1j}) = M_0$ is the saturated value of the magnetization, and $m = 250$ is the chosen number of logarithmically-spaced bins in the $P(1/T_{1})$ distribution. The normalization used to convert $P(1/T_{1})$ into a probability density is detailed in subsection IV B.  See the Supplemental Materials \cite{SuppMat} as well as the references \cite{venkataramanan:ieee2002,song:jmr2002,mitchell:PNMRS2012,singer:jcp2018,singer:jcp2018b, chouzenoux:ieee2010, prange:JMR2009, venkataramanan:jmr2010} for the method to deal with the imperfect inversion in actual experimental data, the details of the mathematical background, and the general procedures for the ILT.   Note that technically, Eq. (\ref{eq:ILTexp}) is the discrete form of a Fredholm integral equation of the first kind \cite{fordham:diff2017}, but for simplicity we refer to it here as an ILT.

The advantage of ILT is that we need not assume a phenomenological functional forms for $M(t)$, such as Eq. (\ref{eq:stretch}) or (\ref{eq:FastSlow}). The only assumption in the ILT is that $M(t)$ decays as a sum of exponentials with decay rates $1/T_{1j}$, implying a heterogeneous distribution in $1/T_1$ over the sample. If the distribution of $1/T_{1}$ is peaked at one value, $P(1/T_{1})$ deduced from $M(t)$ will have one peak (i.e. be a delta function if $1/T_1$ has no distribution) at the most likely value of $1/T_1$. On the other hand, if $1/T_{1}$ is distributed around two distinctive values, as is the case for Eq. (\ref{eq:FastSlow}), $P(1/T_{1})$ will have two peaks centered at $1/T_{1}^{fast}$ and $1/T_{1}^{slow}$.  

\begin{figure}[!ht]
	\begin{center}
		\includegraphics[width=1\columnwidth]{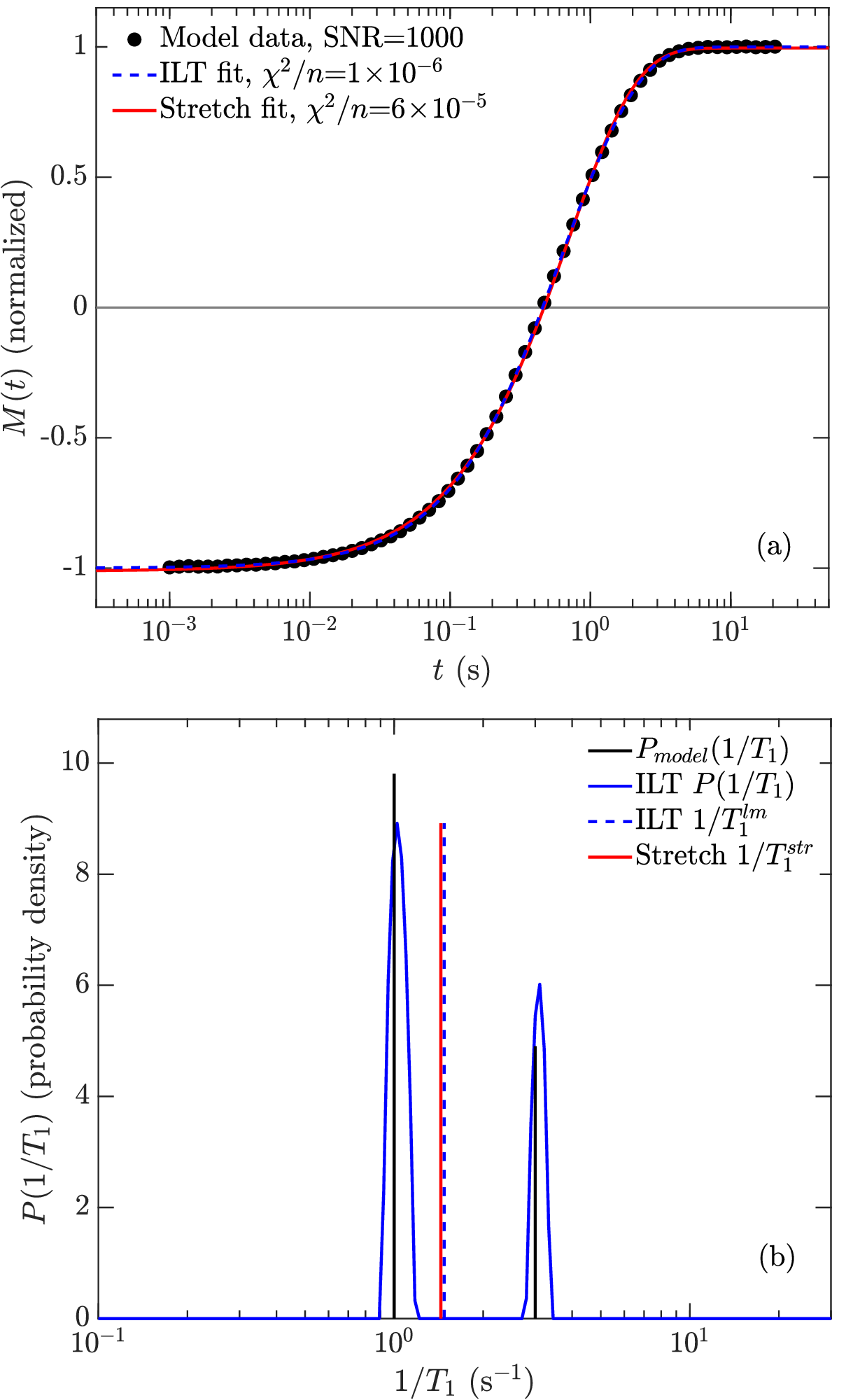}
		\caption{(a) Model data $M(t)$ ($\bullet$), normalized by $M(t)/M_0$, generated for demonstration purposes using Eq. (\ref{eq:model}) as described in the main text, with $n=64$ log-spaced data points. Solid red curve is the best stretched single exponential fit with Eq. (\ref{eq:stretch}) with $1/T_{1}^{str} = 1.44$~s$^{-1}$ and $\beta = 0.89$, under the false assumption that there is only one peak in $P(1/T_1)$ centered around $1/T_{1}^{str}$. Dashed blue curve is the best ILT fit with Eq. (\ref{eq:ILTexp}) resulting in $P(1/T_{1})$, represented by the blue curve in panel (b). Note how the ILT fit yields a lower  mismatch $\chi^2/n$ than the stretched exponential fit. (b) Black vertical lines represent $P_{model}(1/T_{1})$ from Eq. (\ref{eq:model}), consisting of two delta functions at $1/T_{1}^{fast}$ and $1/T_{1}^{slow}$. Blue curve represents the ILT spectrum $P(1/T_{1})$ (probability density) of the $M(t)$ data points in (a). Notice that $1/T_{1}^{str}$ (red vertical line) is a good approximation of the log-mean $1/T_{1}^{lm}$ (dashed blue vertical line) of $P(1/T_{1})$, however the stretched fit misses all of the detail in the underlying model $P(1/T_{1})_{model}$.}
		\label{fig:ILTexample}
	\end{center}
\end{figure}

It is useful to show how ILTT$_{1}$ analysis works based on a simple example. In Fig. \ref{fig:ILTexample}(a), we generated a model relaxation curve $M(t)$ consisting of discrete data points represented by black dots. The best fit of the model data with Eq. (\ref{eq:stretch}) (red curve) yields $1/T_{1}^{str} = 1.44$~s$^{-1}$ and $\beta = 0.89$. The fit seems very good, and the deviation from $\beta=1$ is fairly small. Accordingly, one would be tempted to conclude that $P(1/T_1)$ is narrow and peaked around 1.5~s$^{-1}$, and hence may be close to a delta function as represented by the red vertical line in Fig. \ref{fig:ILTexample}(b). 

However in reality, we generated the discrete $M(t)$ data in Fig. \ref{fig:ILTexample}(a) using a two component function as such:
\begin{equation}
M(t) = 1-2/3 \, e^{-3t} -4/3 \, e^{-t}. \label{eq:model}
\end{equation}   
This is similar to the model in Eq. (\ref{eq:FastSlow}), with $1/T_{1}^{fast} = 3$~s$^{-1}$, $1/T_{1}^{slow} = 1$~s$^{-1}$, $M_{o}=1$, $A_{fast}=2/3$, and $A_{slow}=4/3$, plus random noise at the level of 0.1\% (i.e. a signal to noise ratio of SNR = 1000). The ILT of the model data in Fig. \ref{fig:ILTexample}(a) results in $P(1/T_{1})$ with double peaks at $1/T_{1}^{fast}$ and $1/T_{1}^{slow}$, as shown by a blue curve in Fig. \ref{fig:ILTexample}(b). The log-mean (i.e. center of gravity on a log scale) of $P(1/T_{1})$ is located at $1/T_{1}^{lm}=1.48$~s$^{-1}$ (represented by a dashed blue line). The finite width of the blue curve originates from the discreteness of the model data and the built-in random noise \cite{SuppMat}.   

This simple example illustrates the power of ILT, and the potentially risky nature of the commonly used stretched exponential fit. While $1/T_{1}^{str}$ is indeed close to the real log-mean $1/T_{1}^{lm}$ of $P(1/T_{1})$, the imagined distribution spectrum shown by a single vertical red line in Fig. \ref{fig:ILTexample}(b) does not even remotely resemble the true, double peak structure in $P(1/T_{1})$.  

\section{Experimental}
We grew a single crystal sample of La$_{1.875}$Ba$_{0.125}$CuO$_{4}$ ($T_{c}=4$~K) based on the traveling solvent floating zone technique at Tohoku \cite{Fujita2004}. We aligned and cut a small piece of single crystal with the total mass of 51~mg for this study. We conducted $^{139}$La NMR measurements at 9~T applied along the $c$-axis with the standard $\pi/2 - \pi$ spin-echo pulse sequence. The $^{139}$La NMR lineshapes observed for the nuclear spin $I_{z}=+1/2 \leftrightarrow-1/2$ central transition were very similar to an earlier report \cite{BaekPRB2017}. 

We note that La$_{1.875}$Ba$_{0.125}$CuO$_{4}$ showed a partial loss of $^{139}$La signal (i.e. partial $^{139}$La wipeout) in the vicinity of $\sim$35 K, where the signal dropped to $\sim$1/3 of the full intensity. As such, the ILT may underestimate the distribution in fast $1/T_1$ components in the vicinity of $\sim$35 K. This is similar to previous reports of $^{139}$La wipeout in La$_{1.8-x}$Eu$_{0.2}$Sr$_{x}$CuO$_{4}$ (see Fig. 16 in \cite{HuntPRB2001}), but will not affect our conclusions. By contrast, La$_{1.885}$Sr$_{0.115}$CuO$_{4}$ showed no signs of $^{139}$La wipeout for $T> T_c$ \cite{ArsenaultPRB2019}. 

\begin{figure}
	\begin{center}
		\includegraphics[width=1\columnwidth]{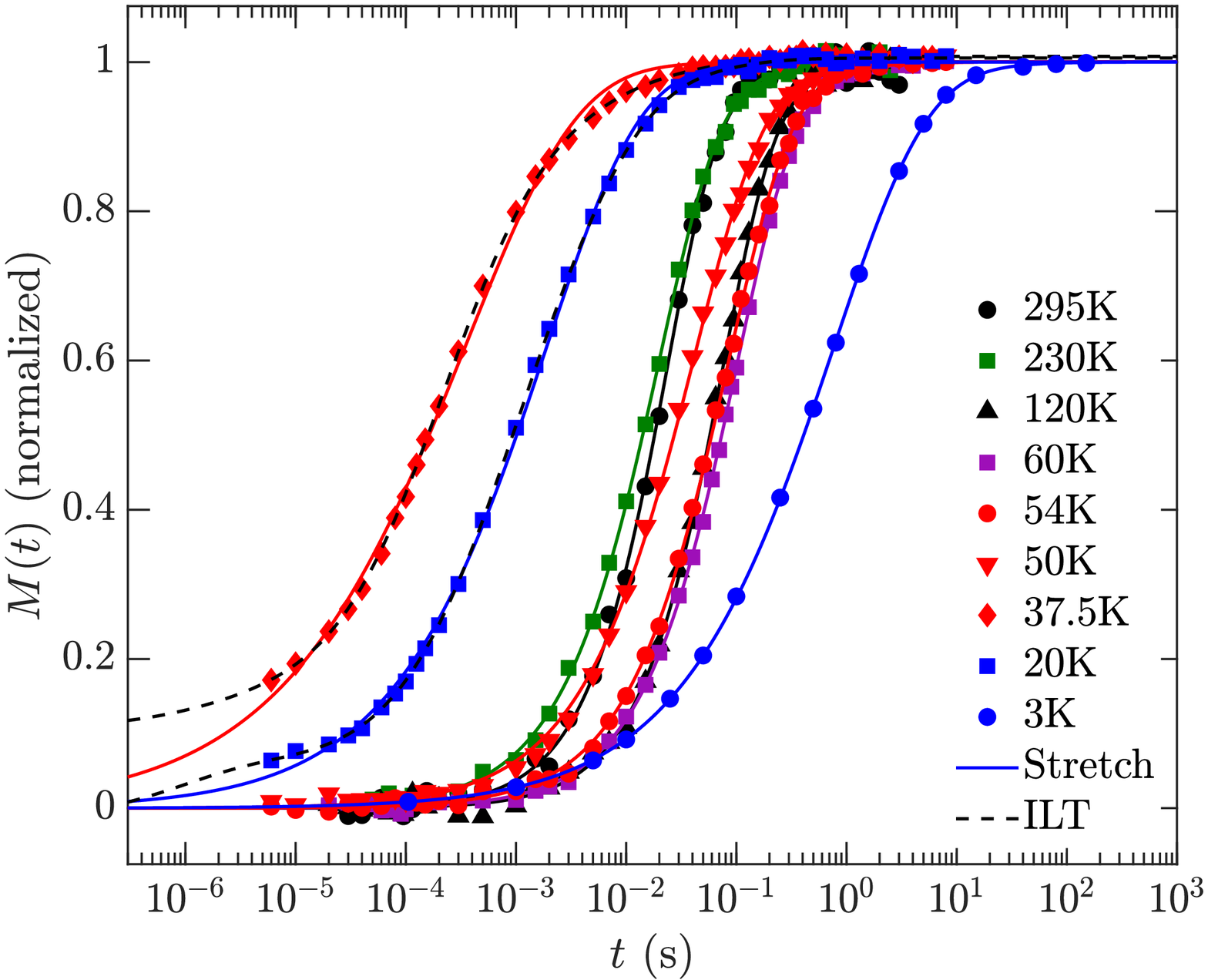}
		\caption{$^{139}$La NMR recovery curves $M(t)$ at representative temperatures, normalized by $(M(t)- M_0)/A+1$, where $M_0$ and $A$ are parameters from the best fits (solid curves) to the stretched exponential form in Eq. (\ref{eq:stretchLa}). Also shown are best fits to the ILT (dashed black lines) using Eq. (\ref{eq:ILTLa}) at 37.5~K and 20~K. Notice that the large distribution in $P(1/T_{1})$ near $T_{spin}^{\mu SR}$ leads to poor stretched exponential fits at 37.5~K and 20~K, while the ILT yields good fits.}
		\label{fig:Mt}
	\end{center}
\end{figure}

We measured $1/T_{1}$ using the inversion recovery method by applying a $\pi$ pulse prior to the spin echo sequence. We summarize the representative results of $M(t)$ in Fig. \ref{fig:Mt}. Measurements of $1/T_{1}^{str}$ with reasonable accuracy is an easy task, and can be usually completed in less than 1 hour at each temperature. However, the accuracy of $M(t)$ required for ILT is far less forgiving, because the resolution of the ILT curve $P(1/T_{1})$ can depend on the experimental noise (i.e. the signal-to-noise ratio) of the $M(t)$ curve \cite{SuppMat}. In addition, we had to use long spin echo recycling time up to 240 s between the spin-echo sequences, so that we can properly capture the longest components of $1/T_{1}$ in $M(t)$. For these reasons, it took up to 24 hours to measure $M(t)$ at a given temperature.  

\section{$^{139}\mathrm{La}$ NMR Results and Discussions}

\subsection{Conventional stretched fit results $1/T_{1}^{str}$}
Before we delve into the ILT analysis of the $M(t)$ data, let us first examine the consequence of the fit with the stretched exponential version of Eq. (\ref{eq:expLa}) commonly used in the literature:
\begin{equation}
M(t) = M_0 - A \,\sum_{k=1}^4 p_k e^{-(q_k t/T_{1}^{str})^{\beta}}, \label{eq:stretchLa}
\end{equation}
where the same normal modes $k$ \cite{Andrew1961, Narath1967} are used as in Eq. \ref{eq:expLa}, and $\sum_{k=1}^4 p_k = 1$ is normalized. We caution that, unlike the case of $I=1/2$, there is no mathematical justification to place the same $\beta$ in all four terms, although we will show below that this phenomenological procedure works fairly well to estimate the log-mean $1/T_1^{lm}$ of the underlying probability density function $P(1/T_1)$.

\begin{figure}
	\begin{center}
		\includegraphics[width=1\columnwidth]{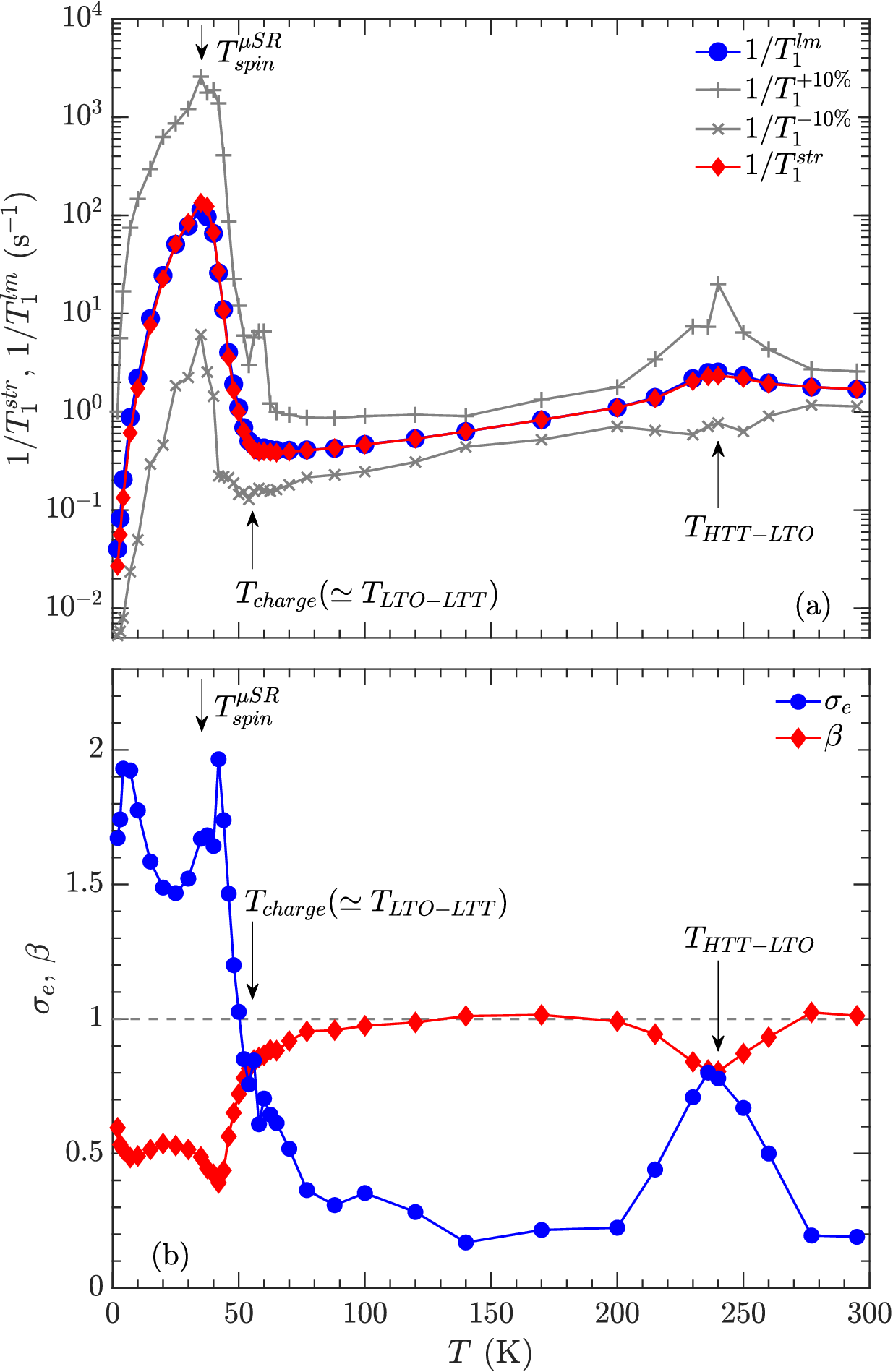}
		\caption{(a) Temperature dependence of $1/T_{1}^{str}$ (red symbols) obtained from the stretched fit with Eq. (\ref{eq:stretchLa}), and log-mean value $1/T_{1}^{lm}$ (blue symbols) of the probability density $P(1/T_{1})$ from Eq. (\ref{eq:T1LM}), along with $1/T_{1}^{+10\%}$ ($1/T_{1}^{-10\%}$) at the top (bottom) 10\% value of the distributed $P(1/T_{1})$, respectively. $1/T_{1}^{str}$ may be considered a close approximation to $1/T_{1}^{lm}$.  See Fig.\ 4 in \cite{ArsenaultPRB2019} for the corresponding results observed for La$_{1.885}$Sr$_{0.115}$CuO$_{4}$.  (b) Temperature dependence of the stretched exponent $\beta$, and standard deviation $\sigma_e$ of $P(1/T_{1})$ from Eq. \ref{eq:sigma}. Note that $\sigma_e$ and $\beta$ are found to be anti-correlated \cite{SuppMat}.}
		\label{fig:LM_sigma}
	\end{center}
\end{figure}

We summarize $1/T_{1}^{str}$ and $\beta$ observed at the $^{139}$La sites in Fig. \ref{fig:LM_sigma}(a) and (b), respectively. $1/T_{1}^{str}$ begins to increase sharply at $T_{charge}$, where charge order turns on a strong enhancement of low-frequency spin fluctuations in the charge ordered domains \cite{TranquadaPRB59, HuntPRL1999}. Our finding is consistent with our powder NQR results by Hunt {\it et al.} \cite{HuntPRB2001} and a more recent single crystal NMR report by Baek {\it et al.} \cite{BaekPRB2017}. $1/T_{1}^{str}$ peaks around $T_{spin}^{\mu SR}$, and is progressively suppressed in the spin ordered state. 

Note that $1/T_{1}^{str}$ is mildly enhanced near $T_{HTT-LTO} \simeq 236$~K, where $\beta$ shows a minimum. This is not due to enhanced spin fluctuations, but rather to the contribution of slow EFG fluctuations near the structural phase transitions. A proof may be found in $^{63}$Cu NMR measurements of $1/T_1$, which show no anomalies at $T_{HTT-LTO}$ \cite{ImaiJPSJ1990, TouLBCOT1, ImaiPRB2019}. $1/T_1$ is three orders of magnitude larger at the $^{63}$Cu sites owing to much a stronger hyperfine coupling with Cu electron spins, and hence less sensitive to such slow EFG fluctuations near the structural phase transition \cite{ImaiJPSJ1990, TouLBCOT1, ImaiPRB2019}. Strictly speaking, one cannot rely on Eq. (\ref{eq:expLa}) or (\ref{eq:stretchLa}) under the presence of EFG contributions to the $T_1$ process. This is because Eq. (\ref{eq:expLa}) or (\ref{eq:stretchLa}) implicitly assume that only the magnetic transitions with $\Delta I_{z} = \pm 1$ between two adjacent nuclear spin energy levels contribute to the $T_1$ process \cite{Andrew1961, Narath1967}, whereas the EFG fluctuations induce $\Delta I_{z} = \pm 2$ transitions as well. In practice, it is difficult to determine the spin and EFG induced $1/T_1$ contributions separately \cite{Suter1998}, and we phenomenologically rely on Eq. (\ref{eq:stretchLa}) to account for the enhanced $1/T_1$ due to the slowly fluctuating EFG.   

Below $\simeq 200$~K down to $\simeq 80$~K, $\beta$ is close to 1. This implies that the $T_1$ process is dominated by Cu spin fluctuations once the HTT-LTO structural phase transition is complete, and the distribution of $1/T_{1}$ is small. In the case of the superconducting compositions above $x \simeq 1/8$, $1/T_{1}$ continues to decrease smoothly toward $T_c$ \cite{Kobayashi1989, Yoshimura1992, BaekPRB2017}. In contrast, $1/T_{1}^{str}$ in the present case begins to level off below $\simeq 80$~K, and $\beta$ deviates from $1$ again. In this temperature range, we expect a growth in nano-scale electronic inhomogeneity as we previously reported for La$_{2-x}$Sr$_{x}$CuO$_{4}$ \cite{SingerPRL2002}. In fact, we recently confirmed that $1/T_{1}$ at $^{63}$Cu sites of the same La$_{1.875}$Ba$_{0.125}$CuO$_{4}$ crystal levels off below $\simeq 80$~K if measured with the pulse separation time $\tau = 2$~$\mu$s between the 90 and 180 degree pulses, whereas $1/T_{1}$ keeps decreasing if measured with $\tau = 20$~$\mu$s \cite{ImaiPRB2019}. As such, our results below $\simeq 80$~K in Fig. \ref{fig:LM_sigma} are consistent with enhanced spin fluctuations with growing spatial distributions. However, what about the potential influence of the dynamic short-range charge order observed above $T_{charge}$ \cite{MiaoPRX2019}, and the slowing of phonons near the LTO to LTT transition toward $T_{charge}$? The slowing fluctuations of charge and lattice would certainly induce slow EFG fluctuations that can potentially contribute to $1/T_1$, as observed near $T_{HTT-LTO}$. We will address this issue below based on ILT in subsection IV C.

\subsection{The ILT results and consistency with the stretched fit}

\begin{figure}
	\begin{center}
		\includegraphics[width=0.85\columnwidth]{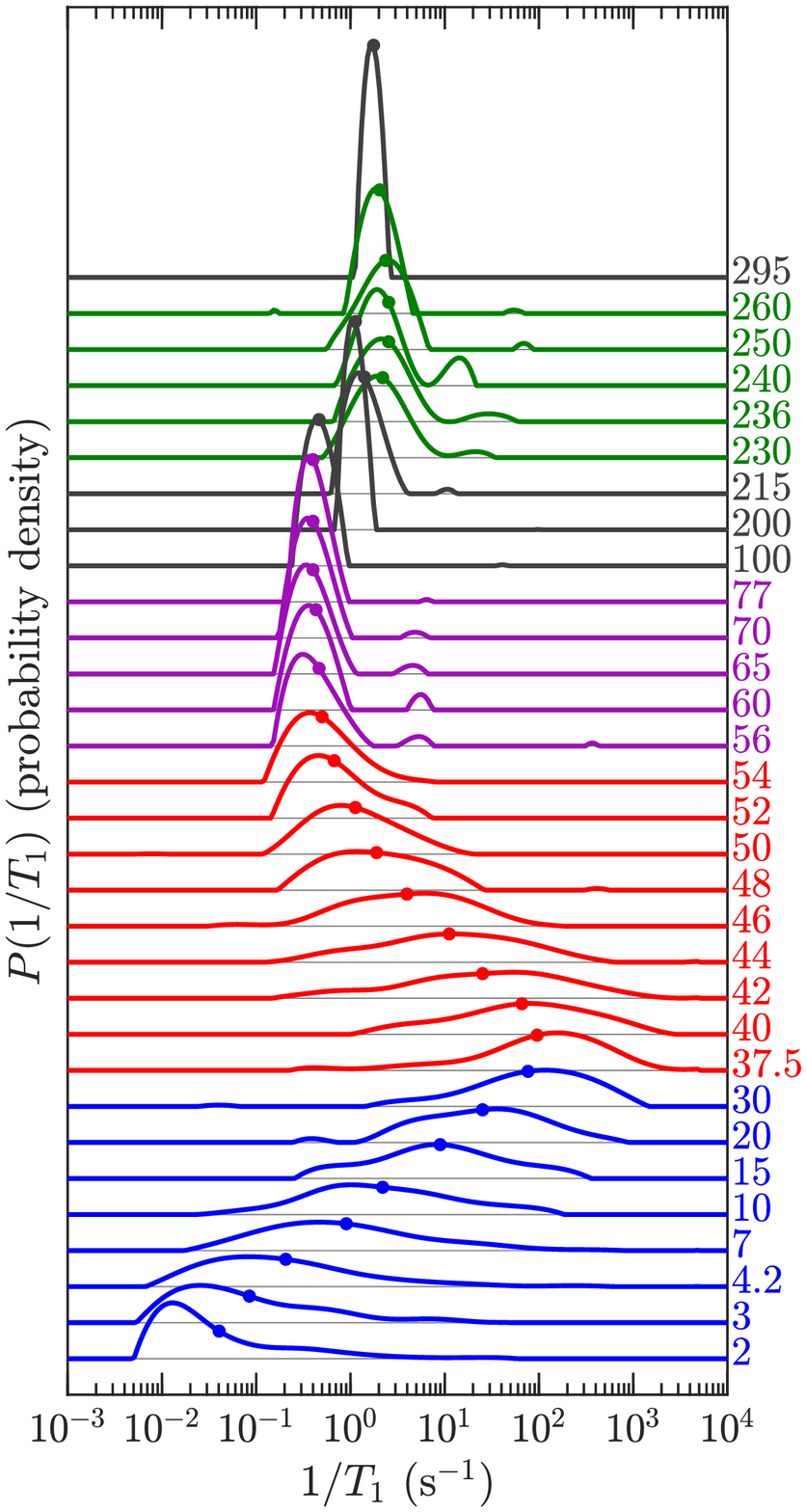}
		\caption{Representative results of $P(1/T_{1})$ (probability density) obtained from $M(t)$ based on ILT. For clarity, the origin of the vertical axes are shifted at each temperature, where the temperature $T$(K) is listed on the right-hand side. The location of the log-means $1/T_1^{lm}$ ($\bullet$) are shown on each $P(1/T_{1})$.}
		\label{fig:T1log}
	\end{center}
\end{figure}

\begin{figure}
	\begin{center}
		\includegraphics[width=1\columnwidth]{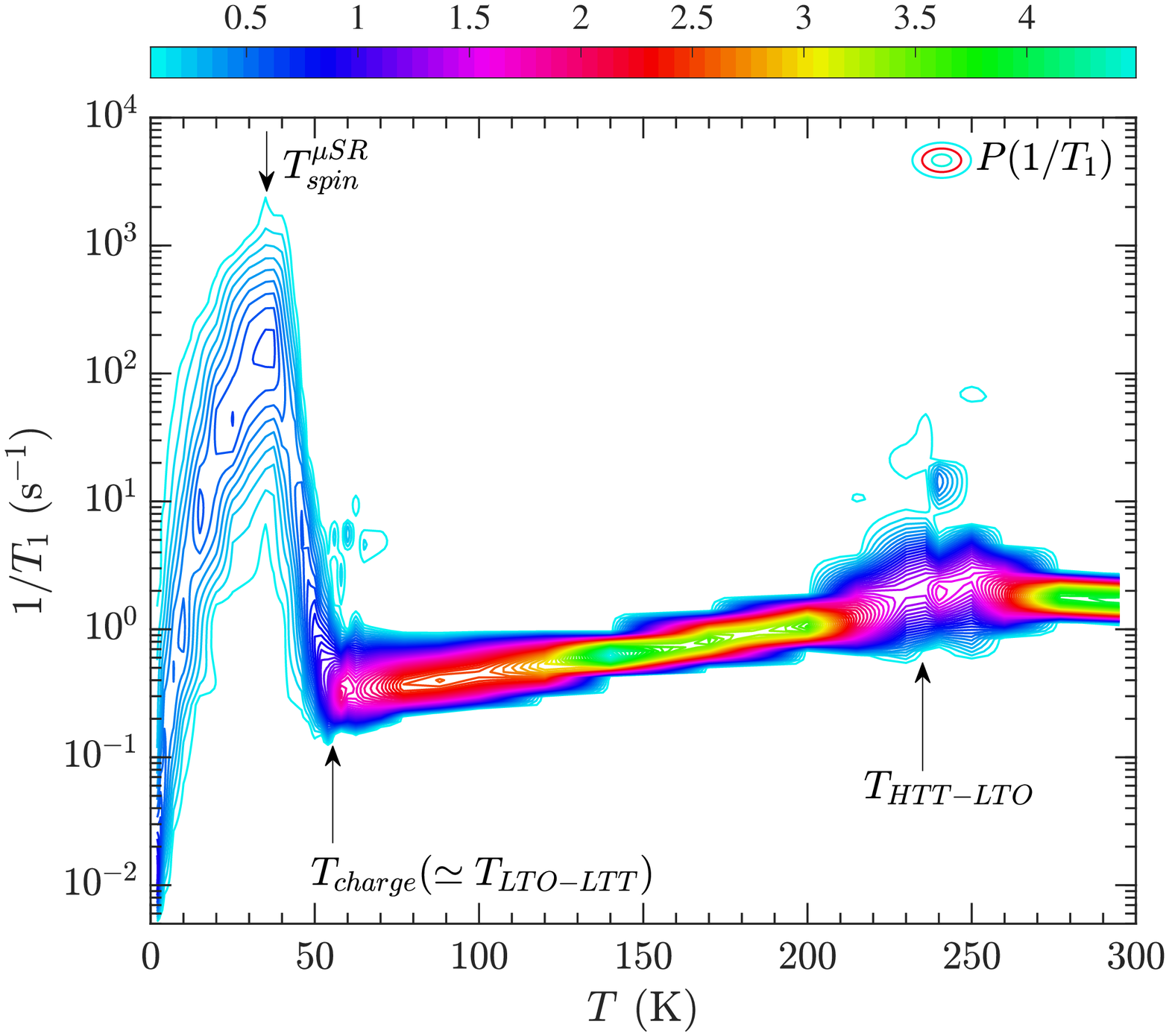}
		\caption{Contour map of $P(1/T_{1})$ (probability density) generated from ILT, using 64 contours. Color bar scale is shown at the top of the figure.}
		\label{fig:T1contour}
	\end{center}
\end{figure}

To account for the 4 normal modes for $I=7/2$ in Eq. (\ref{eq:expLa}), we replaced Eq. (\ref{eq:ILTexp}) with the following:
\begin{align}
M(t) = \sum_{j=1}^{m}\sum_{k=1}^{4}\left[1- 2 p_k e^{-q_k t/T_{1j}}\right]\!P(1/T_{1j}) \label{eq:ILTLa},
\end{align}
where the same normal modes $k$ \cite{Andrew1961, Narath1967} are used as in Eq. (\ref{eq:expLa}), and $\sum_{k=1}^4 p_k = 1$ is normalized. The summation $\sum_{j=1}^m P(1/T_{1j}) = M_0$ is the saturated value of the magnetization, and $m = 250$ is the chosen number of logarithmically-spaced bins ranging from $10^{-3}\, {\rm s^{-1}} \leq1/T_{1j} \leq10^{5} \,{\rm s^{-1}}$. The probability density is then normalized to $\sum_{j=1}^m\!P(1/T_{1j})\,\Delta_P =1$, where the constant $\Delta_P = \log_{10}(1/T_{1j+1}) - \log_{10}(1/T_{1j}) = 0.0321$ is the logarithmic bin spacing. This normalization ensures a 1-1 comparison of $P(1/T_{1})$'s when the bin spacing is different for each $P(1/T_{1})$. Using a $\log_{10}$ base to define $\Delta_P$ conveniently yields unit area for a square $P(1/T_{1})$ distribution a decade wide and of unit height. 

In Fig. \ref{fig:T1log}, we summarize the ILT curves $P(1/T_{1})$ obtained from $M(t)$ at representative temperatures. We also show the evolution of $P(1/T_{1})$ with temperature in the color contour map in Fig. \ref{fig:T1contour}. From each of these ILT curves, we deduced $1/T_{1}^{lm}$ and $\sigma_e$, the log-mean of the distribution $P(1/T_{1})$ and the log-standard deviation of $1/T_{1}$, respectively, as such:  
\begin{align}
\ln(1/T_1^{lm}) = \sum_{j=1}^m \ln(1/T_{1j}) P(1/T_{1j}) \Delta_P &,\label{eq:T1LM} \\
\sigma_e^2 = \sum_{j=1}^m  \left[\ln(1/T_{1j})  - \ln(1/T_1^{lm}) \right]^2 \! P(1/T_{1j})\Delta_P &,\label{eq:sigma}
\end{align}
where $\sum_{j=1}^m\!P(1/T_{1j})\,\Delta_P =1$. We use the subscript $e$ in $\sigma_e$ to emphasize that the logarithm to base $e$ (i.e. the natural logarithm) is used to compute the log-standard deviation. We summarize $1/T_{1}^{lm}$ in comparison to $1/T_{1}^{str}$ in Fig. \ref{fig:LM_sigma}(a), while $\sigma_e$ is compared with $\beta$ in Fig. \ref{fig:LM_sigma}(b). $1/T_{1}^{str}$ and $1/T_{1}^{lm}$ agree well. That is, $1/T_{1}^{str}$ estimated from the phenomenological stretch fit may be considered as a good approximation for the average value of the distributed $1/T_1$. $\sigma_e$ also shows clear anti-correlation with $\beta$. This also makes sense. $\beta$ becomes smaller than 1 when $1/T_{1}$ develops a distribution, whereas $\sigma_e$ increases when the distribution of $1/T_{1}$ grows and $P(1/T_{1})$ becomes wider. Thus we have established that our ILT results encompass the equivalent information as the stretched exponential analysis of $M(t)$. Note however that besides $1/T_{1}^{str}$ and $\beta$, the stretched exponential analysis loses all other information about the underlying probability density $P(1/T_1)$. In the Supplemental Materials \cite{SuppMat} we present details of the ILT analysis, including the concept of ``resolution" and uncertainties in $P(1/T_{1})$.

\subsection{Distribution of $1/T_{1}$}
Having established the validity of the ILT, let us take an additional step and look into exactly how the distribution of $1/T_1$ develops. At the top of Fig. \ref{fig:T1log} is $P(1/T_{1})$ result at 295~K shown in dark gray. $P(1/T_{1})$ is single-peaked, with the log-mean at $1/T_{1}^{lm} = 1.7$~s$^{-1}$. Note that the stretched exponential fit of $M(t)$ returns $\beta =1.01$ at 295~K, and hence the distribution of $1/T_1$ is minimal. The finite width of $P(1/T_{1})$ at 295~K primarily originates from the finite signal-to-noise ratio of the measurement \cite{SuppMat}.

We summarize the $P(1/T_{1})$ curves in the vicinity of $T_{HTT-LTO} \simeq 235$~K in green in Fig. \ref{fig:T1log}. Two noticeable changes take place near $T_{HTT-LTO}$. First, the main peak of $P(1/T_{1})$ shifts to the right, accompanied by significant broadening. The small shift corresponds to the small increase observed for $1/T_{1}^{lm}$ as well as $1/T_{1}^{str}$. The broadening is a consequence of the additional transitions caused by EFG fluctuations, which are not explicitly taken into account in Eq. (\ref{eq:ILTLa}). Second, a small but noticeable split-off peak consistently emerges with $1/T_{1} \simeq 20$~s$^{-1}$. Note that this does not necessarily mean that the EFG induced transition has an order of magnitude faster $1/T_1$ in a small volume fraction of the sample. As noted above, the additional contributions by the fluctuating EFG to the relaxation processes with $\Delta I_{z} = \pm 1$ and $\pm 2$ could significantly modify the relaxation function itself; we take this effect into account only phenomenologically in Eq.(\ref{eq:ILTLa}) which is derived for purely magnetic relaxation. The slow EFG fluctuations cease to exist when the second order structural phase transition is complete, and the split-off peak disappears as we go deeper into the LTO structure below $T_{HTT-LTO}$. $P(1/T_{1})$ regains a narrower, single-peaked structure at 215~K and below down to 100~K, as shown by dark gray curves. 

The ILT results from $\simeq 80$~K down to $T_{charge} \simeq 54$~K are shown in purple. We recall that the charge order transition is accompanied by a nearly first order LTO to LTT structural phase transition \cite{Fujita2004, MiaoPRX2019}.  Interestingly, a split-off peak analogous to that observed around $T_{HTT-LTO}$ emerges again, signaling the presence of slow EFG fluctuations at the NMR frequency scale. In a separate study based on $^{63}$Cu NMR, we show that $^{63}$Cu NMR lineshape exhibits strong magnetic broadening in this temperature range below $\simeq 80$~K prior to the onset of long range charge order at $T_{charge}$. Moreover, $^{63}$Cu $1/T_1$ begins to distribute \cite{ImaiPRB2019}. Taken together, these findings suggest that, regardless of the exact origin of the slow EFG fluctuations above $T_{charge}$ detected here, spin correlations grow hand in hand with the slow lattice and/or charge fluctuations below $\simeq 80$~K.  

We use red curves to show the ILT results in the charge ordered state below $T_{charge} \simeq 54$~K down to the spin ordering temperature $T_{spin}^{\mu SR} \simeq 35$~K. At 54~K, the split-off peak due to EFG fluctuations is suppressed. This is consistent with the nearly first-order nature of the simultaneous LTT and charge order transitions \cite{Fujita2004, MiaoPRX2019}. Once entering the long range charge ordered state, $P(1/T_{1})$ begins to broaden asymmetrically by transferring some spectral weight to larger values of $1/T_1$. At 48~K, approximately 1/3 of the spectral weight still remains at $1/T_{1} \simeq 1$~s$^{-1}$ or below, although the fastest component reaches $1/T_{1} \simeq 30$~s$^{-1}$. In other words, not all the Cu spin fluctuations begin to slow down and enhance $1/T_1$ immediately below $T_{charge}$. Upon further cooling, the entire $P(1/T_{1})$ curve shifts to larger values of $1/T_{1}$ while increasing in width. All $^{139}$La nuclear spins relax with highly enhanced $1/T_{1}$ by 37~K, followed by spin ordering at $T_{spin}^{\mu SR} \simeq 35$~K. A physical picture that emerges from these observations is that {\it the volume fraction of the canonically behaving segments of CuO$_2$ planes with slow $1/T_{1}$ values gradually decreases below $T_{charge}$}, and the spin order at the relatively slow measurement time scale of $\mu$SR experiments sets in when $\sim$100\% volume fraction of the CuO$_2$ planes is under the influence of enhanced spin fluctuations triggered by charge order.   

We summarize the ILT results below $T_{spin}^{\mu SR}$ using blue curves.  The ILT curve $P(1/T_{1})$ progressively shifts its weight to lower values of $1/T_{1}$ as the fluctuating spins freeze toward the base temperature.  By 7~K, a majority of $^{139}$La sites relax with $1/T_{1} \simeq 1$~s$^{-1}$ or slower. It is in this temperature range where Zeeman perturbed $^{63}$Cu NQR signals become observable with increasing intensity \cite{TouLBCOT2, HuntPRB2001}, since the hyperfine magnetic field from frozen Cu moments become static at the NMR measurement time scale. In a separate work, we will demonstrate that a cut off introduced for $P(1/T_{1})$ can naturally account for the fraction of the observable $^{63}$Cu NMR signal intensity that arises from canonically behaving domains below $T_{charge}$, {\it i.e.} $^{63}$Cu signal intensity wipe out effects \cite{ImaiPRB2019}.

\subsection{Comparison with La$_{1.885}$Sr$_{0.115}$CuO$_{4}$}
\begin{figure}
	\begin{center}
		\includegraphics[width=1\columnwidth]{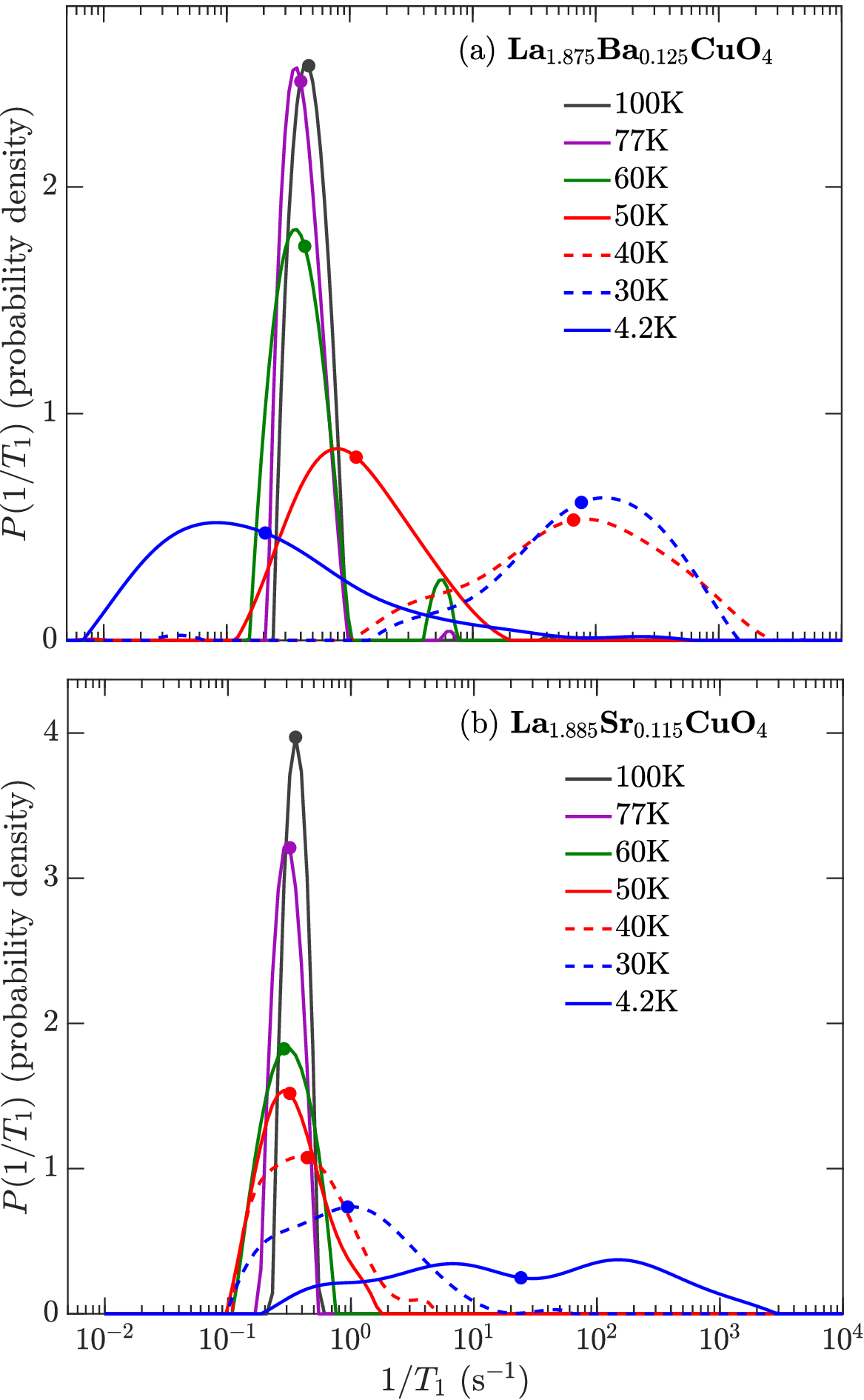}
		\caption{Comparison of the $P(1/T_{1})$ (probability density) between (a) La$_{1.875}$Ba$_{0.125}$CuO$_{4}$ (this work, $T_{charge} \simeq 54$~K, $T_{c}=4$~K) and, (b) La$_{1.885}$Sr$_{0.115}$CuO$_{4}$ (adopted from \cite{ArsenaultPRB2018}, $T_{charge} \simeq 80$~K, $T_{c}=31$~K), at selected temperatures. The location of the log-means $1/T_1^{lm}$ ($\bullet$) are shown on each $P(1/T_{1})$.}
		\label{fig:LSCO}
	\end{center}
\end{figure}

In Fig. \ref{fig:LSCO}, we compare the $P(1/T_{1})$ results for La$_{1.875}$Ba$_{0.125}$CuO$_{4}$ with our earlier report for La$_{1.885}$Sr$_{0.115}$CuO$_{4}$ \cite{ArsenaultPRB2019}. $P(1/T_{1})$ for La$_{1.885}$Sr$_{0.115}$CuO$_{4}$ begins to broaden below the onset of its charge order at $T_{charge} \simeq 80$~K \cite{Croft, Thampy, WenNatComm2019}, without exhibiting the split-off peak arising from EFG fluctuations. Since no LTT structural phase transition exists in La$_{1.885}$Sr$_{0.115}$CuO$_{4}$, this might be an indication that the signature of the EFG fluctuations observed above $T_{charge}$ in the present case of La$_{1.875}$Ba$_{0.125}$CuO$_{4}$ is due primarily to the slow lattice fluctuations rather than slow charge fluctuations. On the other hand, the Bragg peaks associated with charge order in La$_{1.885}$Sr$_{0.115}$CuO$_{4}$ are known to be very weak, and hence charge fluctuations may also be too weak to induce the split-off peak in $P(1/T_{1})$.

Another interesting aspect is the qualitative difference in the way the distribution of $P(1/T_{1})$ develops. In the case of La$_{1.885}$Sr$_{0.115}$CuO$_{4}$, notice that $P(1/T_{1})$ continues to transfer its spectral weight to smaller values down to $1/T_1 \simeq 0.1$ s$^{-1}$ even below $T_{charge} \simeq 80$~K, although the log-mean of the distribution is shifting to larger values of $1/T_1$. This means that $1/T_{1}$ continues to become smaller at a substantial fraction of $^{139}$La sites even deep inside the charge ordered state below $T_{charge} \simeq 80$~K. The temperature dependence of $1/T_{1}$ at these $^{139}$La sites is qualitatively the same as that in the canonical superconducting CuO$_2$ planes in La$_{1.85}$Sr$_{0.15}$CuO$_{4}$ \cite{Kobayashi1989, Yoshimura1992, BaekPRB2017}. It is just that the {\it volume fraction} of such canonically superconducting domains gradually decreases below $T_{charge}$. As a consequence, we can clearly see that the peak structure from the canonically behaving domains is still visible as a clearly identifiable shoulder at $1/T_1 \simeq 0.1$ s$^{-1}$ down to $\sim 30$~K. This conclusion was corroborated by the fact that two types of $^{63}$Cu NMR signals exist in La$_{1.885}$Sr$_{0.115}$CuO$_{4}$, too: the wing like signal is extremely broad and with large $1/T_{1}$, whereas the canonically behaving signal has a narrow lineshape with slow $1/T_{1}$ that is comparable to the optimally doped superconductor with $x=0.15$ \cite{ImaiPRB2017}.

In contrast, the slow shoulder at $1/T_1 \simeq 0.1$ s$^{-1}$ observed for La$_{1.885}$Sr$_{0.115}$CuO$_{4}$ is not observable in the present case of La$_{1.875}$Ba$_{0.125}$CuO$_{4}$. As temperature decreases below $T_{charge} \simeq 54$~K, $P(1/T_{1})$ shifts to larger values of $1/T_{1}$ more quickly. There is only a hint of a slow shoulder at $1/T_{1}\simeq 3$~$s^{-1}$ from 54~K down to 48~K in Fig. \ref{fig:T1log}.  We confirmed that the $^{63}$Cu NMR lineshape for La$_{1.875}$Ba$_{0.125}$CuO$_{4}$ broadens more homogeneously below $\simeq 80$~K, and lacks the aforementioned two component wing plus narrow-peak structure observed for La$_{1.885}$Sr$_{0.115}$CuO$_{4}$ \cite{ImaiPRB2019}. In short, CuO$_2$ planes in La$_{1.875}$Ba$_{0.125}$CuO$_{4}$ are more homogeneously affected by charge order and accompanying enhancements of slow spin fluctuations, which increases $1/T_1$.

\section{Summary and conclusions}
We have reported the ILTT$_{1}$ analysis to demonstrate how the distribution of $1/T_1$ develops in charge ordered La$_{1.875}$Ba$_{0.125}$CuO$_{4}$. We identified the signature of the slow EFG fluctuations near $T_{HTT-LTO}$, and showed that the same signature reemerges above $T_{charge}$. Our experiments and ILTT$_{1}$ analysis cannot determine whether the source of the slow EFG fluctuations is the lattice and/or charge vibrations.  

By comparing the probability density function $P(1/T_1)$ from ILTT$_{1}$ analysis, we demonstrated that the magnetic properties of the CuO$_2$ planes become highly inhomogeneous below $T_{charge} \simeq 54$~K, and domains with canonical behavior of a high $T_c$ superconductor persist even below $T_{charge}$.  These residual domains are oblivious to the charge order transition, and spin fluctuations are not anomalously enhanced.  This suggests that charge order does not set in homogeneously at $T_{charge}$ in the entire CuO$_{2}$ planes.  This finding is consistent with the fact that the width of the charge order Bragg peaks ({\it i.e.} the inverse of the charge order correlation length) is not resolution limited above $\simeq 40$~K, and hence the size of the charge ordered domains is not infinite \cite{MiaoPRX2019}.

The volume fraction of these residual domains gradually diminishes below $T_{charge}\simeq 54$~K in La$_{1.875}$Ba$_{0.125}$CuO$_{4}$.  By 40~K, nearly $\simeq 100$\% volume of the CuO$_2$ planes have large $1/T_{1}$ induced by enhanced low frequency Cu spin fluctuations triggered by charge order.  This finding is in stark contrast with the case of La$_{1.885}$Sr$_{0.115}$CuO$_{4}$, where a significant fraction of domains in the CuO$_2$ planes still exhibits characteristic behavior of high $T_c$ superconductor with slowing $1/T_{1}$ at 40~K \cite{ArsenaultPRB2018, ArsenaultPRB2019}. These contrasting behaviors are consistent with the fact that superconductivity sets in at as high as $T_{c} = 31$~K for La$_{1.885}$Sr$_{0.115}$CuO$_{4}$, while superconductivity is strongly suppressed to $T_{c} =4$~K in La$_{1.875}$Ba$_{0.125}$CuO$_{4}$.  

The present work also highlights the usefulness of the ILTT$_{1}$ analysis technique in general. The ILT provides us with much richer information than the conventional stretched fit analysis since ILT generates the probability density function $P(1/T_1)$ rather than just the average value of $1/T_1$. Moreover, it is important to note that ILTT$_{1}$ analysis is unbiased, and one does not need to assume the shape of the distribution function $P(1/T_1)$. The novel ILTT$_{1}$ analysis, which has been used successfully in NMR petrophysics, has the potential to revolutionize NMR research of quantum materials with disorder.   

Finally, we briefly comment on some earlier publications, in which several groups tried to model the distribution of $1/T_{1}$ in cuprates \cite{HuntPRB2001, Curro, Mitrovic} and unrelated materials \cite{Gezo, Dioguardi}. Note that all these earlier attempts were made by assuming a functional form of $P(1/T_1)$ up front, which may or may not reflect the reality. For example, our earlier attempt in 2001 assumed a symmetrical Gaussian distribution of $1/T_1$ on a log scale in the charge ordered state of cuprates for fitting $M(t)$, as shown in Fig.\ 12 of \cite{HuntPRB2001}. The present work shows that such a symmetrical functional form is only approximately true for La$_{1.875}$Ba$_{0.125}$CuO$_{4}$ in a limited temperature range, and invalid for La$_{1.885}$Sr$_{0.115}$CuO$_{4}$ \cite{ArsenaultPRB2019}. Ref. \cite{Curro} assumed instead that magnetic inhomogeneity of the CuO$_2$ planes in Eu co-doped 214 cuprates is caused entirely by preexisting quenched disorder, which gives rise to a Gaussian distribution in an activation energy for spin fluctuations rather than the distribution of $1/T_1$ itself. Since their toy model is based on an incorrect assumption that charge order transition does not exist, all the NMR properties are expected to evolve smoothly, in contradiction with the experimental reality such as Fig. \ref{fig:LM_sigma}(a). Ref. \cite{Mitrovic} extended our earlier analysis in \cite{HuntPRB2001}, and assumed a symmetrical distribution function for $1/T_{1}$ on a log scale for La$_{1.88}$Sr$_{0.12}$CuO$_{4}$ \cite{Mitrovic}. Despite the unrealistic assumption in contradiction to the experimental reality of non-symmetric $P(1/T_{1})$ \cite{ArsenaultPRB2019}, their analysis actually showed the onset of unusual NMR anomalies starting from $\simeq 80$~K. However, the authors in Ref. \cite{Mitrovic} did not attribute their findings to the onset of charge order, as they had been advocating for the absence of charge order in the superconducting phase of La$_{2-x}$Sr$_{x}$CuO$_{4}$. 

\begin{acknowledgments}
The authors thank J. Wang for helpful discussions. P.M.S. is supported by The Rice University Consortium for Processes in Porous Media. The work at McMaster is supported by NSERC. The work at Tohoku is supported by Grant-in-Aid for Scientific Research (A) (16H02125), Japan.

\end{acknowledgments}

\newpage
\noindent
{\bf Supplemental Materials}

\section{Inverse Laplace Transform}

The inverse Laplace transform (ILT) consists of fitting the measured inversion recovery curve $M(t)$ to a sum of exponentials with decay rate $1/T_{1j}$ and weight $P(1/T_{1j}) \geq 0$. The advantage of ILT over stretched exponential fits is that the ILT does not assume any phenomenological functional forms for the decay in $M(t)$. The only assumption in the ILT is that $M(t)$ decays as a sum of exponentials with decay rates $1/T_{1j}$, implying a heterogeneous distribution in $1/T_{1}$ over the sample. Put in other words, $P(1/T_{1})$ is the probability density function for $1/T_{1}$ over the heterogeneous sample.

In the case of $^{139}$La nuclear spin with $I=7/2$, we need to take into account four normal modes in the $1/T_{1}$ relaxation process \cite{Andrew1961, Narath1967}.  The discrete form of the ILT inverts for $P(1/T_{1j})$ assuming the following expression for the experimental data $M(t)$ measured at a set of time $t_{i}$:
\begin{equation}
M(t_i) = \sum_{j=1}^{m}\sum_{k=1}^{4}\left[1- 2 p_k e^{-q_k t_i/T_{1j}}\right]\!P(1/T_{1j}) + E(t_i).
\label{eq:ILTsum}
\end{equation}
The coefficients \{$p_k$,$q_k$\} (with $k = 1,..,4$) are calculated for each of the aforementioned normal modes \cite{Andrew1961, Narath1967} as such $p_k = \{1/84,3/44,75/364,1225/1716\}$ and $q_k = \{1,6,15,28\}$, and $\sum_{k=1}^4 p_k = 1$ is normalized. $E(t)$ is Gaussian noise with standard deviation $\sigma_E$. Technically, Eq. \ref{eq:ILTsum} is the discrete form of a Fredholm integral equation of the first kind \cite{fordham:diff2017}, but for simplicity we refer to it as an ILT. 

The summation $\sum_{j=1}^m P(1/T_{1j}) = M_0$ in Eq. \ref{eq:ILTsum} is the saturated value of the magnetization, and $m $ is the chosen number of logarithmically-spaced bins in the $P(1/T_{1})$ distribution. The probability density is then normalized to $\sum_{j=1}^m\!P(1/T_{1j})\,\Delta_P =1$, where the constant $\Delta_P = \log_{10}(1/T_{1j+1}) - \log_{10}(1/T_{1j})$ is the logarithmic bin spacing. This normalization ensures a 1-1 comparison of $P(1/T_{1})$'s when the bin spacing is different for each $P(1/T_{1})$. Using a $\log_{10}$ base to define $\Delta_P$ conveniently yields unit area for a square $P(1/T_{1})$ distribution a decade wide and of unit height. 

The column vector form for Eq. \ref{eq:ILTsum} is given by:
\begin{equation}
{\bf M} = K \, {\bf P} +{\bf E}, \,\,\,\,{\bf P}\geq 0, \label{eq:ILTvec}
\end{equation}
where the following are defined:
\begin{align}
{\bf M} &= \left\{M(t_i),i = 1,....,n\right\},\nonumber \\
K &= \left\{K_{ij} = \sum_{k=1}^4\left[1- 2 p_k e^{-q_k t_i/T_{1j}}\right]\right\}, \label{eq:ILTdef}\\
{\bf P} &= \left\{P(1/T_{1j}),j = 1,....,m\right\},\nonumber
\end{align}
where $K$ is the kernel matrix of size $n\times m$, and {\bf E} is the noise vector. The number of measured data points in the present case is between $n = 18 \leftrightarrow 46$, and $m$ is chosen to be $m = 250$. For computational efficiency, the kernel $K$ is compressed using singular value decomposition (SVD) \cite{venkataramanan:ieee2002,song:jmr2002,mitchell:PNMRS2012}, which is not detailed here in the interest of brevity. However, given the relatively small size of the $K$ matrix ($n\times m  \simeq 46\times250$ typically), there is no need for data compression in the present case.
 
The disadvantage of ILT is that it is an ill-posed problem in the sense that for a given set of data with finite noise $E$, many solutions will fit the data within the statistics of the noise. Various solutions exist to deal with such ill-posed problems \cite{mitchell:PNMRS2012}, including Tikhonov regularization \cite{butler:SIAM1981,venkataramanan:ieee2002,song:jmr2002,singer:jcp2018,singer:jcp2018b}, maximum entropy \cite{chouzenoux:ieee2010}, Monte Carlo \cite{prange:JMR2009}, and the Mellin transform \cite{venkataramanan:jmr2010}. 

In this report we choose the Tikhonov regularization method given the extensive literature on the subject. Regarding uncertainties in the optimal solution {\bf P}, we note that there are no uncertainties or ``error bars" associated with each bin $j$ of the $P(1/T_{1j})$ distribution. Instead, uncertainties in the moments of {\bf P} such as $1/T_1^{lm}$ (log-mean) or $\sigma_e$ (log standard deviation) can be quantified. Furthermore, uncertainties in the total area of {\bf P} (i.e. $\sum_{j=1}^m P(1/T_{1j}) = M_0$) and the partial area of {\bf P} below (i.e. $\sum_{j=1}^{m,cut} P(1/T_{1j})$) or above (i.e. $\sum_{j={m,cut}}^{m} P(1/T_{1j})$) a certain a cutoff $(1/T_{1})_{cut}$ can also be quantified. Details of how to quantify these uncertainties can be found in \cite{prange:JMR2009,venkataramanan:jmr2010}, which are beyond the scope in this report.

\subsection{Non-negative optimization}

Using the Tikhonov regularization method, the goal is to find the solution {\bf P} which minimizes the cost function \cite{butler:SIAM1981}:
\begin{equation}
{\bf P} = \underset{{\bf P}\geq0}{\mathrm{arg\, min}}\,\,  || {\bf M} - K \,{\bf P}||^2 + \alpha ||{\bf P}||^2   \label{eq:ILTarg}
\end{equation}
by using non-negative least-squares, where $||..||$ is the vector norm. The first term is the residual between data and fit, and the second term is the regularization factor. $\alpha$ is the scalar regularization parameter (i.e. a smoothing factor) chosen to be large enough to make the solution stable in the presence of noise. The solution {\bf P}($\alpha$) in Eq. \ref{eq:ILTarg} implicitly depends on $\alpha$. Choosing $\alpha = 0$ leads to a unique solution {\bf P}($\alpha = 0$), however the solution is ``spiky", where the position and amplitude of the spikes depend on the particular noise realization {\bf E}; this undesirable case is so called ``under-regularized". In the opposite extreme, choosing a large $\alpha \gg 10$ over-smooths the solution {\bf P}($\alpha \gg 10$); this undesirable case is so called ``over-regularized".

The first task in finding the optimal solution {\bf P} in Eq. \ref{eq:ILTarg} is to choose a reasonable range of $1/T_{1j}$ bins for $P(1/T_{1j})$. Given the large distribution in $P(1/T_{1j})$ at low temperatures, and given the range of times $10^{-5}\, {\rm s}\lesssim  t_i \lesssim10^{2} \,{\rm s}$ in $M(t_i)$, we selected $m$ = 250 bins equally spaced on a logarithmic scale ranging from $10^{-3}\, {\rm s^{-1}} \leq1/T_{1j} \leq10^{5} \,{\rm s^{-1}}$. This results in a logarithmic bin spacing of $\Delta_P = \log_{10}(1/T_{1j+1}) - \log_{10}(1/T_{1j})= 0.0321$. Note that using a log-spaced distribution function $P(1/T_{1j})$ implies that the equivalent linear-spaced distribution function is given by $P_{lin}(1/T_{1j}) = P(1/T_{1j})/(1/T_{1j})$. 

Details of the steps required to optimize the non-negative solution {\bf P} in Eq. \ref{eq:ILTarg} can be found in \cite{butler:SIAM1981,venkataramanan:ieee2002}. The basic steps for the 1D case without SVD decomposition are shown here for convenience. The function $f({\bf c})$ to be minimized with respect to the vector {\bf c} is defined as:
\begin{align}
f({\bf c}) = \frac{1}{2} {\bf c'} \left[G({\bf c}) + \alpha I\right]{\bf c} - {\bf c'}{\bf M},\label{eq:DelQ3}
\end{align}
where the quasi-Newton method \cite{lawson:1974} can be used, with an initial guess of {\bf c} = {\bf 1}, and ${\bf c'}$ represents the transpose of {\bf c}. The {\bf c} vector is the same size as {\bf M}. $I$ is the identity matrix of size {\bf M}$\times${\bf M}. The ``$G$ matrix" of size {\bf M}$\times${\bf M} is defined as:
\begin{align}
G({\bf c}) = K \begin{bmatrix} 
    H\!\left(K'_1 {\bf c}\right) & 0 & \dots & 0 \\
    0 & H\!\left(K'_2 {\bf c}\right)  & \dots & 0 \\
    \vdots & \vdots &  & \vdots \\
    0 &    0  & \dots  & H\!\left(K'_m {\bf c}\right)  
    \end{bmatrix} K'  \label{eq:DelQ2}
\end{align}
where $H$ is the heavy-side function, and $K_j$ is the $j$th column of $K$. 

With the optimal {\bf c} now found, the optimal non-negative solution {\bf P} is given by:
\begin{align}
{\bf P} = \max\left\{0,K'{\bf c}\right\},
 \label{eq:DelQ4}
\end{align}
which is a function of $\alpha$. The next step is to determine the optimal regularization parameter $\alpha_{opt}$ to be used.

\subsection{Optimal regularization parameter}

The residual $\chi(\alpha)$ between data and optimal fit is given by:
\begin{equation}
\chi(\alpha) = \alpha || {\bf c}||= || {\bf M} - K \,{\bf P}(\alpha)||. \label{eq:ILTchi}
\end{equation}
There are then two criteria for establishing $\alpha_{opt}$:
\begin{align}
\chi(\alpha_1)  &=  \sigma_E \sqrt{n},\nonumber \\
\frac{d\ln \!\chi(\alpha) }{d\ln\!\alpha}\bigg|_{\alpha_2} &=\, 10^{-1}, \label{eq:ILTbrd}\\
\alpha_{opt} &= \max \{\alpha_1,\alpha_2 \}. \nonumber
\end{align}

 \begin{figure}[h!]
 	\begin{center}
 		\includegraphics[width=1\columnwidth]{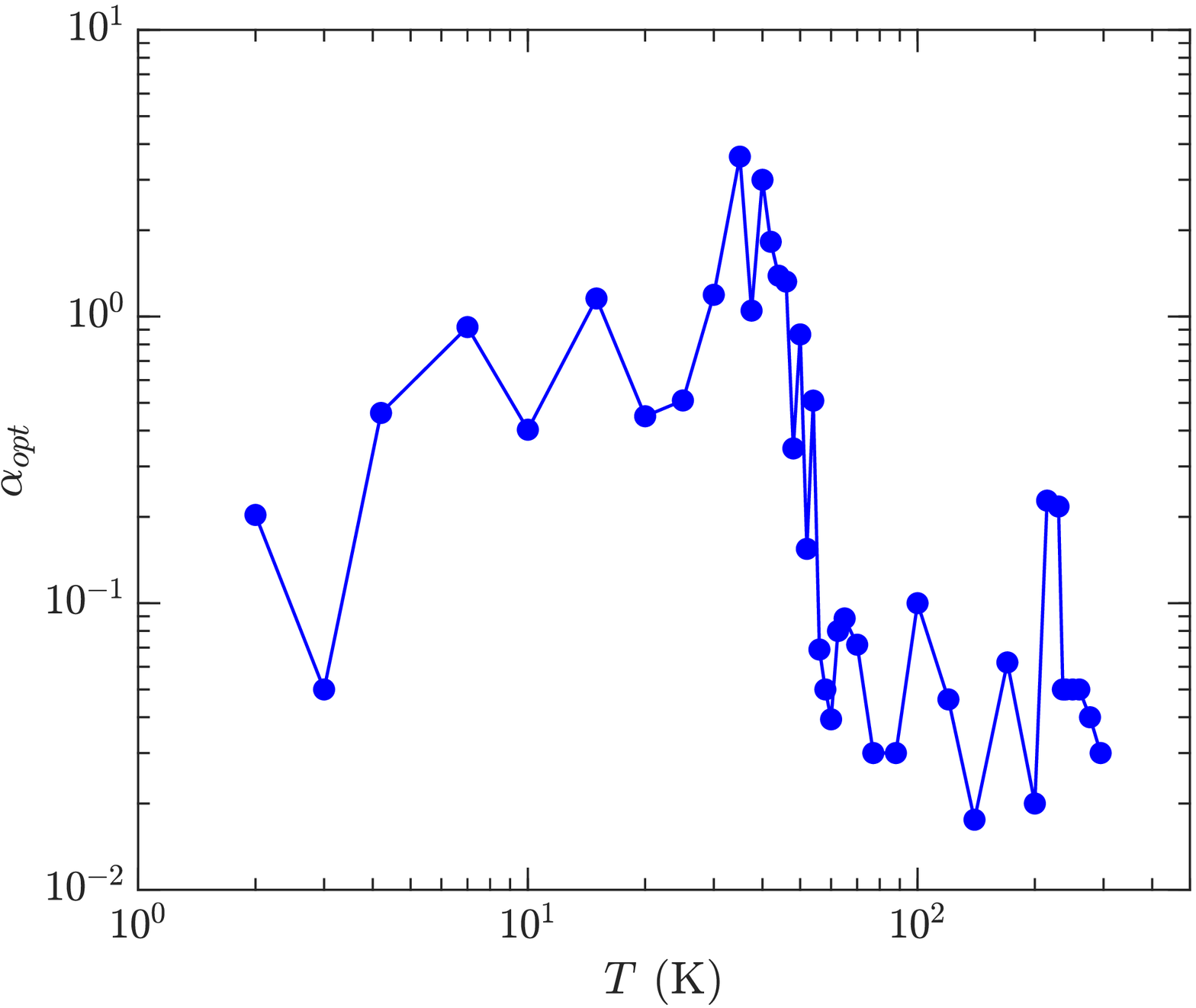}	
 	\end{center}
 	\caption{Temperature dependence of the optimal regularization parameter $\alpha_{opt}$.}\label{fg:Alpha}
 \end{figure}

The first criterion is the Bulter-Reeds-Dawson (BRD) condition \cite{butler:SIAM1981}, where $\alpha_1$ is found such that $\chi(\alpha_1)$ equals the noise $\sigma_E$ times the square-root of the number of data points $\sqrt{n}$ (or the square-root of the number of singular values $\sqrt{n_{\mbox{\tiny SVD}}}$ if SVD is used). The experimental noise $\sigma_E$ can be determined using one of the following techniques: (1) acquiring $M(t)$ data with the RF power turned off, then taking the standard deviation of the data, (2) using the standard deviation of the imaginary channel in $M(t)$, although there may be ringing artifacts, or (3) using SVD and taking the standard deviation of the difference between the data and the projection of the data onto the range space \cite{venkataramanan:ieee2002}.

The BRD condition works well provided $M(t)$ is dominated by statistical noise $E(t)$. However, if systematic errors occur in $M(t)$, then the BRD condition breaks down. Systematic errors occur due to hardware limitations which are only apparent at high SNR, or the kernel $K$ may not be exact due to the presence of quadrupole relaxation. In such cases, a second criterion is introduced known as the ``heel" condition \cite{song:jmr2002}, where $\alpha_2$ is defined when the derivative $d\ln\!\chi(\alpha) /d\ln\!\alpha$ is equal to $10^{-1}$. A derivative of $10^{-1}$ corresponds roughly to the heel of the $\chi(\alpha)$ versus $\alpha $ function, where $d\ln\!\chi(\alpha) /d\ln\!\alpha > 0$ for $\alpha \geq \alpha_{opt}$.

The optimal alpha $\alpha_{opt}$ is then obtained from the maximum of both conditions $\max \{\alpha_1,\alpha_2 \}$. In practice, Eq. \ref{eq:ILTarg} is solved for a range of selected values $\alpha = 10^6 \rightarrow 10^{-3}$ in descending half-decade (or finer) steps, with the misfit $\chi(\alpha)$ and slope $d\ln\!\chi(\alpha) /d\ln\!\alpha$ calculated at each descending step. In order to save computational time, the $\alpha$ descent can be stopped once either $\alpha_1$ or $\alpha_2$ has been reached. Another way to save computation time during the $\alpha$ descent is to use the optimal {\bf c} vector from the previous $\alpha$ computation as the initial guess of the next $\alpha$ computation in Eq. \ref{eq:DelQ3}. 

Fig. \ref{fg:Alpha} shows the temperature dependence of $\alpha_{opt}$ for the dataset, which indicate that $\alpha_{opt}$ increases somewhat at temperatures below $\lesssim 50$ K. This is a result of the heel condition ($\alpha_{opt} = \alpha_2$) being met more often than the BRD condition ($\alpha_{opt} = \alpha_1$).

\subsection{Incomplete inversion at short times}

\begin{figure}[ht!]
	\begin{center}
		\includegraphics[width=1\columnwidth]{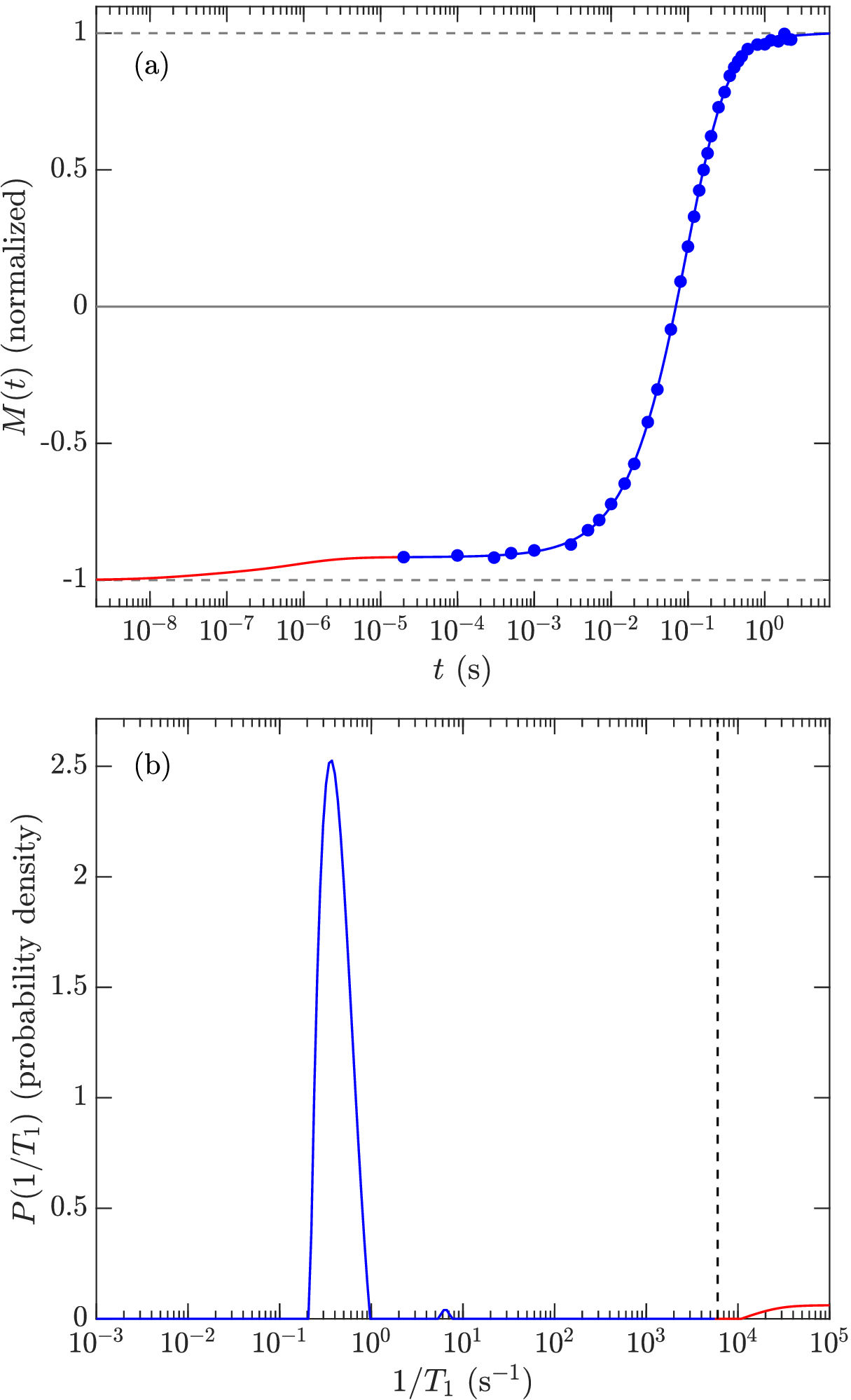}	
	\end{center}
	\caption{(a) Magnetization data $M(t)$ ($\bullet$) and ILT fit (solid curve) at 77 K, where red region ($t < t_{min}$) shows incomplete inversion. (b) Corresponding ILT distribution $P(1/T_{1})$ (solid curve) at 77 K, where red region indicates unphysical region $1/T_1 > (1/T_{1})_{cut}$ (dashed line) used to account for incomplete inversion in $M(t)$.}\label{fg:R1cut}
\end{figure}  

Due to experimental limitations, the inversion recovery following the 180$^{\rm o}$ r.f. pulse is not exact. One way to take this into account is to replace the factor of 2 in Eq. \ref{eq:ILTsum} with a free parameter whose optimized value ends up less than 2. 

A more robust method for accounting for incomplete inversion is to use the $P(1/T_{1})$ distribution itself. The selection of bins $10^{-3}\, {\rm s^{-1}} \leq1/T_{1j} \leq10^{5} \,{\rm s^{-1}}$ in $P(1/T_{1j})$ is chosen to have an extended range at the fast end $(1/T_{1})_{max} = 10^{5} \,{\rm s^{-1}}$. These fast bins simulate fast relaxation components that decay well before the first data point at $t_{min} \simeq 10^{-5}\, {\rm s}$, specifically $\sum_k p_k e^{-q_k t_{min} (1/T_{1})_{max}} \simeq 0.005$. While these fast $1/T_{1}$ components are not physically measurable, they mimic the effect of incomplete inversion of the magnetization at $M(t_{min} \simeq 10^{-5}\, {\rm s})$. 

An example of the incomplete inversion in $M(t)$ is shown in Fig. \ref{fg:R1cut}(a) for 77 K, where the red part of the curve ($t < t_{min}$) shows the incomplete inversion. The corresponding $P(1/T_{1})$ is shown in Fig. \ref{fg:R1cut}(b), where the red part of the curve shows the region $1/T_1 > (1/T_{1})_{cut}$, where $(1/T_{1})_{cut} = 6000 \, {\rm s}^{-1}$ is chosen. The region $P(1/T_{1} > (1/T_{1})_{cut})$ is a fraction $\simeq 0.04$ of the total $P(1/T_{1})$ at 77 K, which accounts for the incomplete inversion of $M(t_{min})/M_0 \simeq -0.92$. $(1/T_{1})_{cut}$ is chosen for each measurement, i.e. at each temperature, based on the extent of the fast region. 

\begin{figure}[h!]
	\begin{center}
		\includegraphics[width=1\columnwidth]{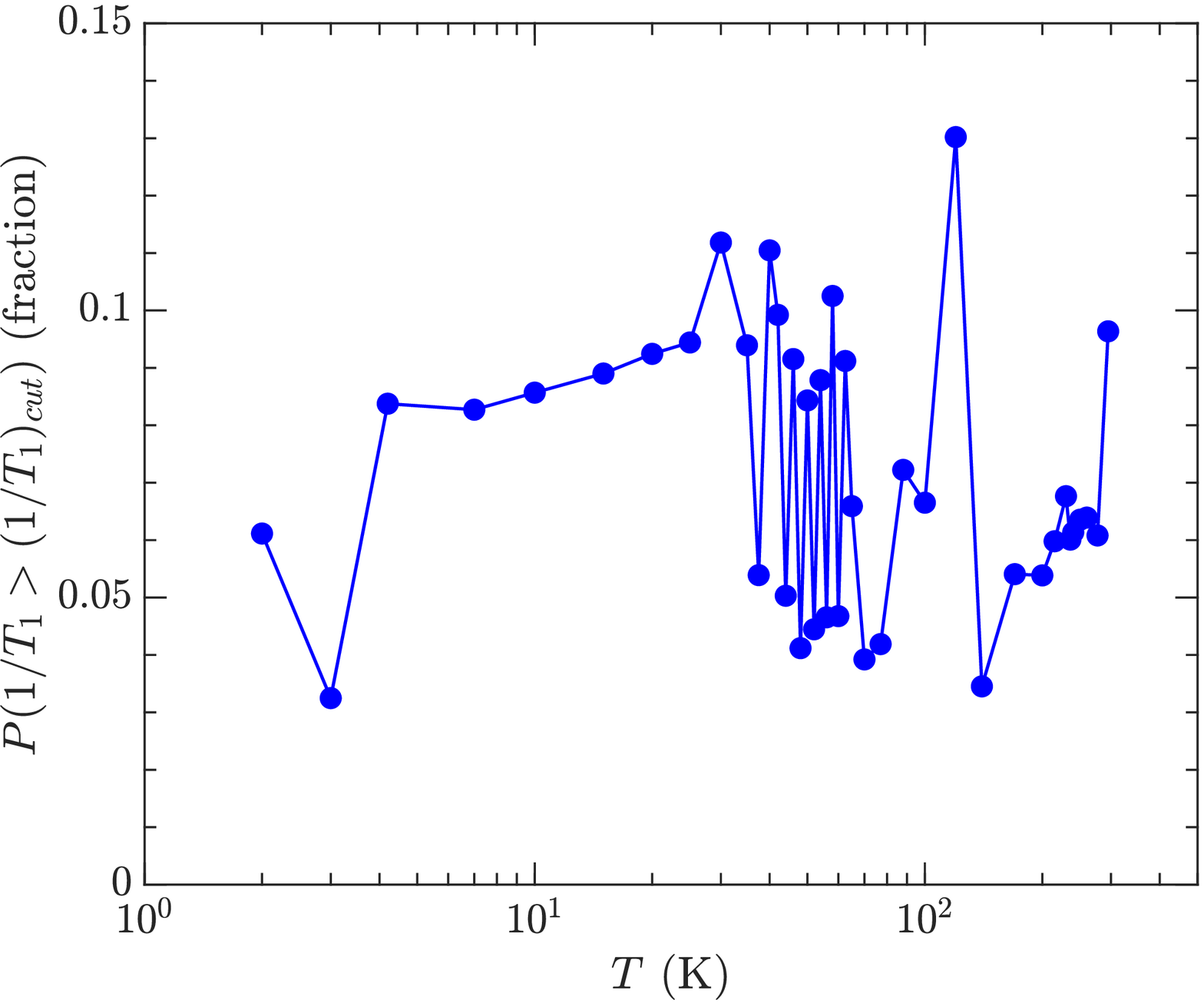}
	\end{center}
	\caption{Temperature dependence of the fraction of signal above the cutoff $P(1/T_{1} > (1/T_{1})_{cut})$ due to incomplete inversion recovery.}\label{fg:Cut}
\end{figure}

Figure \ref{fg:Cut} shows the temperature dependence of the fraction of signal $P(1/T_{1} > (1/T_{1})_{cut})$. We do not see a systematic variation in $P(1/T_{1} > (1/T_{1})_{cut})$ with temperature, in particular it does not correlate with the partial $^{139}$La wipeout temperature in the vicinity of $\sim$35 K. Rather, $P(1/T_{1} > (1/T_{1})_{cut})$ is a function of various acquisition parameters and experimental conditions.


The final step is to truncate the data in the unphysical region $P(1/T_{1j} > (1/T_{1})_{cut})$ by setting those bins to zero, i.e. $P(1/T_{1j} > (1/T_{1})_{cut}) = 0$. In other words, the effect of incomplete inversion is not used in the subsequent analysis of $P(1/T_{1})$.

\subsection{Resolution as a function of SNR}

\begin{figure}[h!]
	\begin{center}
		\includegraphics[width=1\columnwidth]{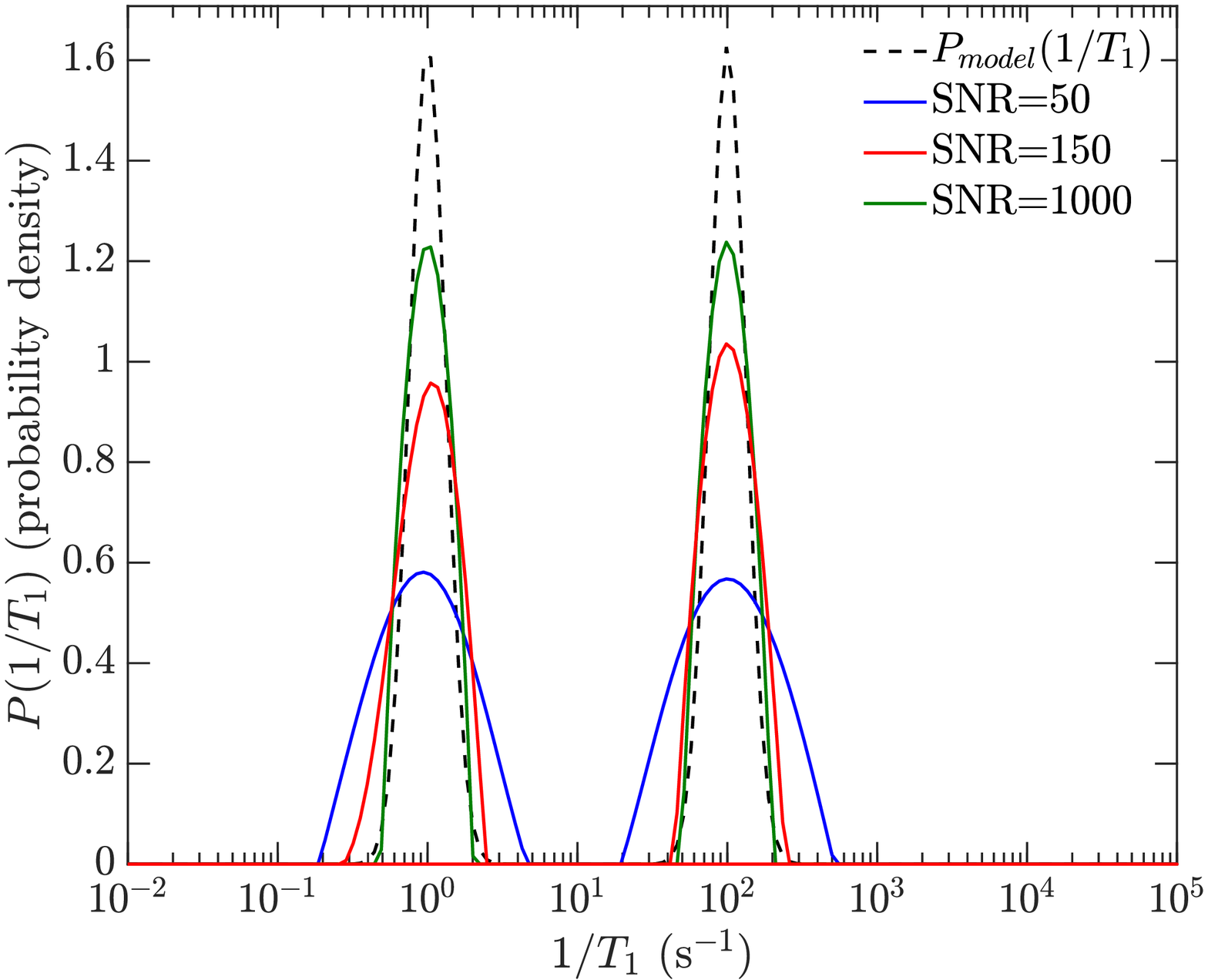}
	\end{center}
	\caption{Forward model $P_{model}(1/T_{1})$, and inverted distributions as a function of SNR. The optimal values for $\alpha_{opt}$ are $\alpha_{opt}$ = $\{0.9,0.03,0.006\}$ for SNR = $\{50,150,1000\}$, respectively.}\label{fg:FEst_SNR}
\end{figure}

\begin{figure}[h!]
	\begin{center}
		\includegraphics[width=1\columnwidth]{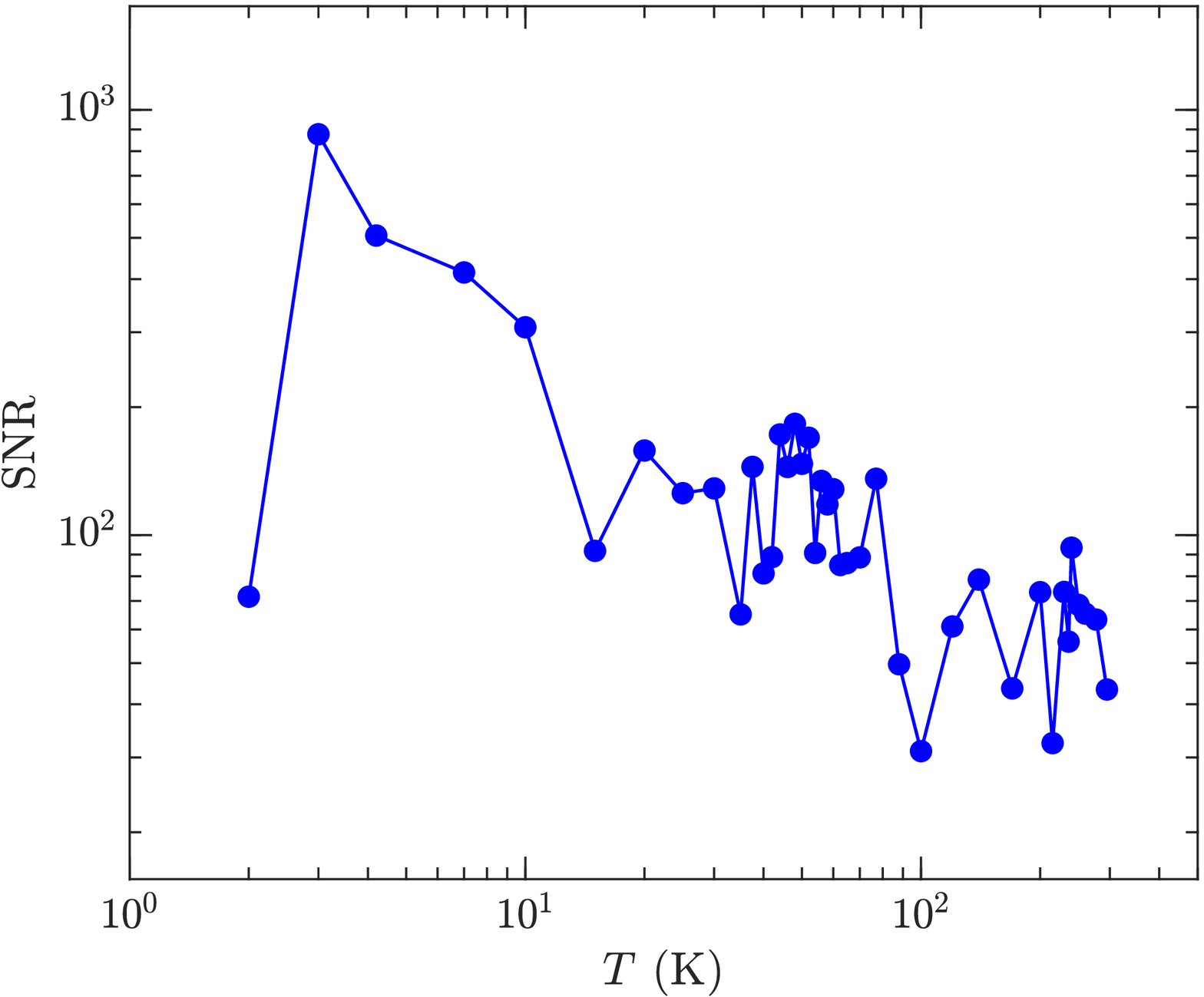}
	\end{center}
	\caption{Temperature dependence of the SNR for the experimental data.}\label{fg:SNR}
\end{figure}

It is helpful to get a sense of the ``resolution'' in $P(1/T_{1})$, as a function of the signal to noise ratio (SNR) in $M(t)$. For this purpose, a forward model $P_{model}(1/T_{1j})$ is generated comprising two log-normal peaks at $1/T_{1j} = 1 \,{\rm s^{-1}}$ and $1/T_{1j} = 10^2 \,{\rm s^{-1}}$ with equal amplitudes and equal widths. The $M(t)$ data are then generated using Eq. \ref{eq:ILTsum} with synthetic Gaussian noise $\sigma_E$. The above inversion is then carried out using three different noise values $\sigma_E$, or equivalently, three different SNR's defined as:
\begin{equation}
{\rm SNR} = \frac{M_0}{\sigma_E}. \label{eq:ILTsnr}
\end{equation}
The resulting distributions as a function of SNR are shown in Fig. \ref{fg:FEst_SNR}, where the selected values SNR = $\{50,150,1000\}$ correspond roughly to the lowest, average, and highest SNR values of the experimental data (see Fig. \ref{fg:SNR}), respectively.

Fig. \ref{fg:FEst_SNR} indicates that the resolution in the inverted distributions increases with increasing SNR, as expected. In other words, the inverted distributions tend more towards $P_{model}(1/T_{1})$ at higher SNR, which is a result of the lower $\alpha_{opt}$ found from the BRD criterion in Eq. \ref{eq:ILTbrd} (i.e. lower $\sigma_E$). The forward model example illustrates the capability of the ILT in separating peaks in $P(1/T_{1})$ as a function of SNR. In the case of the lowest SNR = 50, the peaks are distinguishable provided they are a half-decade apart in $1/T_1$. As such, a half-decade in $1/T_1$ may loosely be considered as the ILT ``resolution" at SNR = 50, although this is only semi-quantitative. 

As shown in Fig. \ref{fg:FEst_SNR}, the resolution can in principle improve with increasing SNR. Fig. \ref{fg:SNR} shows the temperature dependence of the SNR for the experimental data, which indicates that a resolution of less than a half-decade in $1/T_1$ is possible below $\lesssim $ 50 K, provided the BRD condition ($\alpha_{opt} = \alpha_1$) is met rather than the heel condition ($\alpha_{opt} = \alpha_2$). If the heel condition is met, then increasing the SNR does not necessarily improve the resolution.

\clearpage
\section{Comparison between ILT and stretched exponential}

\begin{figure}[ht!]
	\begin{center}
		\includegraphics[width=1\columnwidth]{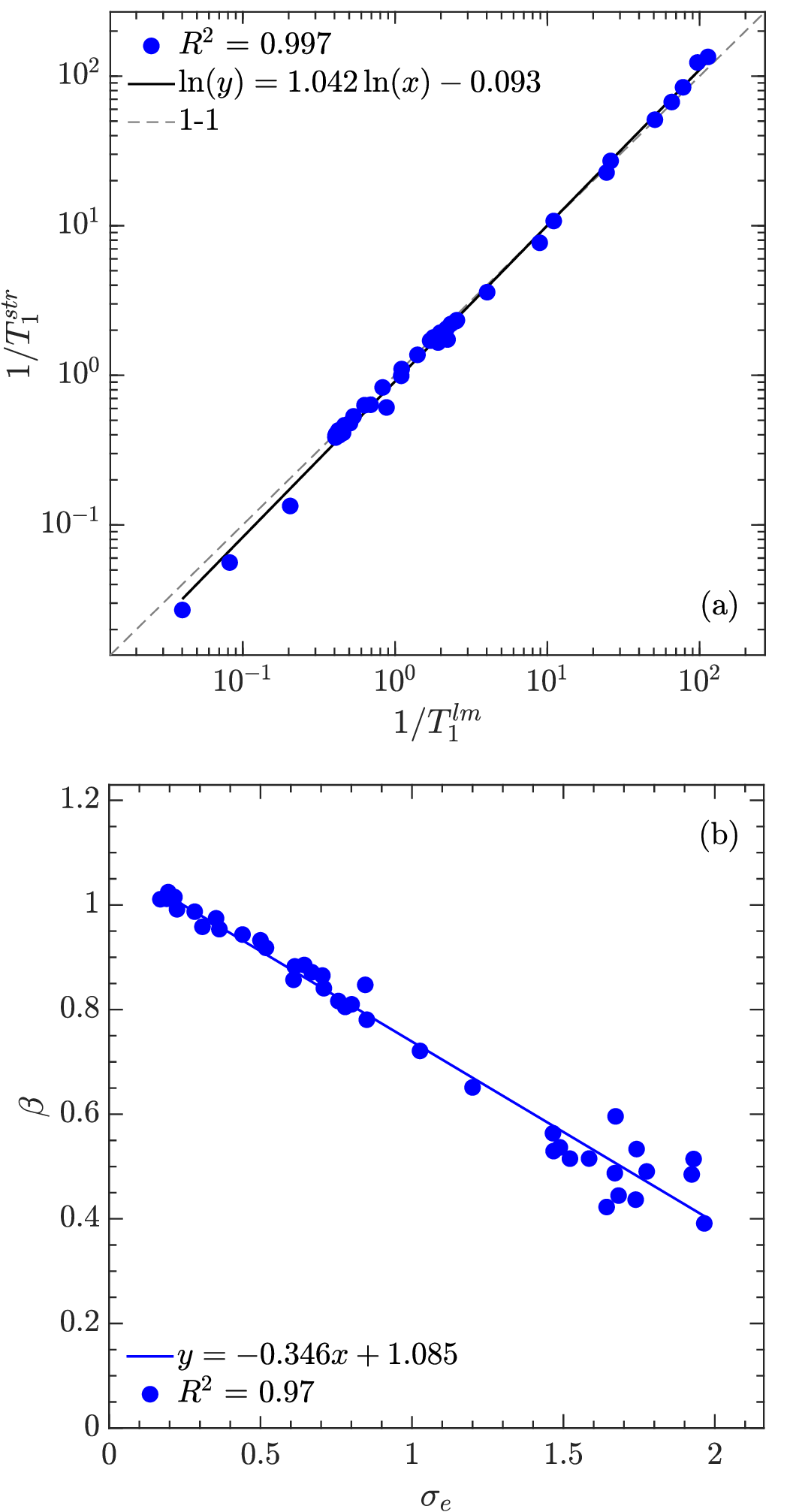}	
	\end{center}
	\caption{(a) Cross plot of log-mean $1/T_{1}^{lm}$ from ILT versus $1/T_{1}^{str}$ from stretched fit. (b) Cross plot of standard deviation $\sigma_e$ from ILT versus stretched exponent $\beta$ from stretched fit.}\label{fg:SigmaCross}
\end{figure}   

Fig. \ref{fg:SigmaCross} shows a cross plot the ILT results versus the stretched exponential results. As shown in Fig. \ref{fg:SigmaCross}(a), a very strong correlation ($R^2 = 0.997$) is found between the log-mean $1/T_{1}^{lm}$ and the stretched $1/T_{1}^{str}$. While in Fig. \ref{fg:SigmaCross}(b), a strong anti-correlation ($R^2 = 0.970$) is found between the log-mean standard-deviation $\sigma_e$ and the stretched exponent $\beta$. Note however that besides $1/T_{1}^{str}$ and $\beta$, the stretched exponential analysis loses all other information about the underlying probability density $P(1/T_1)$. Furthermore, when $P(1/T_1)$ is wide, i.e. when $\beta$ is small ($\beta\lesssim 0.6$), the stretched exponential fit does not necessarily give a good fit to $M(t)$, which results in scattering of $\sigma_e$ versus $\beta$ in Fig. \ref{fg:SigmaCross}(b).



\begin{thebibliography}{54}%
\makeatletter
\providecommand \@ifxundefined [1]{%
 \@ifx{#1\undefined}
}%
\providecommand \@ifnum [1]{%
 \ifnum #1\expandafter \@firstoftwo
 \else \expandafter \@secondoftwo
 \fi
}%
\providecommand \@ifx [1]{%
 \ifx #1\expandafter \@firstoftwo
 \else \expandafter \@secondoftwo
 \fi
}%
\providecommand \natexlab [1]{#1}%
\providecommand \enquote  [1]{``#1''}%
\providecommand \bibnamefont  [1]{#1}%
\providecommand \bibfnamefont [1]{#1}%
\providecommand \citenamefont [1]{#1}%
\providecommand \href@noop [0]{\@secondoftwo}%
\providecommand \href [0]{\begingroup \@sanitize@url \@href}%
\providecommand \@href[1]{\@@startlink{#1}\@@href}%
\providecommand \@@href[1]{\endgroup#1\@@endlink}%
\providecommand \@sanitize@url [0]{\catcode `\\12\catcode `\$12\catcode
  `\&12\catcode `\#12\catcode `\^12\catcode `\_12\catcode `\%12\relax}%
\providecommand \@@startlink[1]{}%
\providecommand \@@endlink[0]{}%
\providecommand \url  [0]{\begingroup\@sanitize@url \@url }%
\providecommand \@url [1]{\endgroup\@href {#1}{\urlprefix }}%
\providecommand \urlprefix  [0]{URL }%
\providecommand \Eprint [0]{\href }%
\providecommand \doibase [0]{https://doi.org/}%
\providecommand \selectlanguage [0]{\@gobble}%
\providecommand \bibinfo  [0]{\@secondoftwo}%
\providecommand \bibfield  [0]{\@secondoftwo}%
\providecommand \translation [1]{[#1]}%
\providecommand \BibitemOpen [0]{}%
\providecommand \bibitemStop [0]{}%
\providecommand \bibitemNoStop [0]{.\EOS\space}%
\providecommand \EOS [0]{\spacefactor3000\relax}%
\providecommand \BibitemShut  [1]{\csname bibitem#1\endcsname}%
\let\auto@bib@innerbib\@empty
\bibitem [{\citenamefont {Axe}\ \emph {et~al.}(1989)\citenamefont {Axe},
  \citenamefont {Moudden}, \citenamefont {Hohlwein}, \citenamefont {Cox},
  \citenamefont {Mohanty}, \citenamefont {Moodenbaugh},\ and\ \citenamefont
  {Xu}}]{Axe1989}%
  \BibitemOpen
  \bibfield  {author} {\bibinfo {author} {\bibfnamefont {J.~D.}\ \bibnamefont
  {Axe}}, \bibinfo {author} {\bibfnamefont {A.~H.}\ \bibnamefont {Moudden}},
  \bibinfo {author} {\bibfnamefont {D.}~\bibnamefont {Hohlwein}}, \bibinfo
  {author} {\bibfnamefont {D.~E.}\ \bibnamefont {Cox}}, \bibinfo {author}
  {\bibfnamefont {K.~M.}\ \bibnamefont {Mohanty}}, \bibinfo {author}
  {\bibfnamefont {A.~R.}\ \bibnamefont {Moodenbaugh}},\ and\ \bibinfo {author}
  {\bibfnamefont {Y.}~\bibnamefont {Xu}},\ }\bibfield  {title} {\bibinfo
  {title} {{Structural phase transformations and superconductivity in
  ${\mathrm{La}}_{2\mathrm{\ensuremath{-}}\mathrm{x}}$${\mathrm{Ba}}_{\mathrm{x}}$${\mathrm{CuO}}_{4}$}},\
  }\href {https://doi.org/10.1103/PhysRevLett.62.2751} {\bibfield  {journal}
  {\bibinfo  {journal} {Phys. Rev. Lett.}\ }\textbf {\bibinfo {volume} {62}},\
  \bibinfo {pages} {2751} (\bibinfo {year} {1989})}\BibitemShut {NoStop}%
\bibitem [{\citenamefont {Luke}\ \emph {et~al.}(1991)\citenamefont {Luke},
  \citenamefont {Le}, \citenamefont {Strenlieb}, \citenamefont {Wu},
  \citenamefont {Uemura}, \citenamefont {Brewer},\ and\ \citenamefont
  {Riseman}}]{Luke1991}%
  \BibitemOpen
  \bibfield  {author} {\bibinfo {author} {\bibfnamefont {G.~M.}\ \bibnamefont
  {Luke}}, \bibinfo {author} {\bibfnamefont {L.~P.}\ \bibnamefont {Le}},
  \bibinfo {author} {\bibfnamefont {B.~J.}\ \bibnamefont {Strenlieb}}, \bibinfo
  {author} {\bibfnamefont {W.~D.}\ \bibnamefont {Wu}}, \bibinfo {author}
  {\bibfnamefont {Y.~J.}\ \bibnamefont {Uemura}}, \bibinfo {author}
  {\bibfnamefont {J.~H.}\ \bibnamefont {Brewer}},\ and\ \bibinfo {author}
  {\bibnamefont {Riseman}},\ }\bibfield  {title} {\bibinfo {title} {Static
  magnetic order in
  {${\mathrm{La}}_{1.875}$${\mathrm{Ba}}_{\mathrm{0.125}}$${\mathrm{CuO}}_{4}$}},\
  }\href@noop {} {\bibfield  {journal} {\bibinfo  {journal} {Physica C}\
  }\textbf {\bibinfo {volume} {185-189}},\ \bibinfo {pages} {1175} (\bibinfo
  {year} {1991})}\BibitemShut {NoStop}%
\bibitem [{\citenamefont {Tranquada}\ \emph {et~al.}(1995)\citenamefont
  {Tranquada}, \citenamefont {Sternlieb}, \citenamefont {Axe}, \citenamefont
  {Nakamura},\ and\ \citenamefont {Uchida}}]{Tranquada1995}%
  \BibitemOpen
  \bibfield  {author} {\bibinfo {author} {\bibfnamefont {J.~M.}\ \bibnamefont
  {Tranquada}}, \bibinfo {author} {\bibfnamefont {B.~J.}\ \bibnamefont
  {Sternlieb}}, \bibinfo {author} {\bibfnamefont {J.~D.}\ \bibnamefont {Axe}},
  \bibinfo {author} {\bibfnamefont {Y.}~\bibnamefont {Nakamura}},\ and\
  \bibinfo {author} {\bibfnamefont {S.}~\bibnamefont {Uchida}},\ }\bibfield
  {title} {\bibinfo {title} {Evidence for stripe correlations in copper oxide
  superconductors},\ }\href@noop {} {\bibfield  {journal} {\bibinfo  {journal}
  {Nature}\ }\textbf {\bibinfo {volume} {375}},\ \bibinfo {pages} {561}
  (\bibinfo {year} {1995})}\BibitemShut {NoStop}%
\bibitem [{\citenamefont {Hunt}\ \emph {et~al.}(1999)\citenamefont {Hunt},
  \citenamefont {Singer}, \citenamefont {Thurber},\ and\ \citenamefont
  {Imai}}]{HuntPRL1999}%
  \BibitemOpen
  \bibfield  {author} {\bibinfo {author} {\bibfnamefont {A.~W.}\ \bibnamefont
  {Hunt}}, \bibinfo {author} {\bibfnamefont {P.~M.}\ \bibnamefont {Singer}},
  \bibinfo {author} {\bibfnamefont {K.~R.}\ \bibnamefont {Thurber}},\ and\
  \bibinfo {author} {\bibfnamefont {T.}~\bibnamefont {Imai}},\ }\bibfield
  {title} {\bibinfo {title} {{$^{63}\mathrm{Cu}$ NQR measurement of stripe
  order parameter in
  ${\mathrm{La}}_{2\ensuremath{-}\mathit{x}}{\mathrm{Sr}}_{\mathit{x}}{\mathrm{CuO}}_{4}$}},\
  }\href {https://doi.org/10.1103/PhysRevLett.82.4300} {\bibfield  {journal}
  {\bibinfo  {journal} {Phys. Rev. Lett.}\ }\textbf {\bibinfo {volume} {82}},\
  \bibinfo {pages} {4300} (\bibinfo {year} {1999})}\BibitemShut {NoStop}%
\bibitem [{\citenamefont {Hunt}\ \emph {et~al.}(2001)\citenamefont {Hunt},
  \citenamefont {Singer}, \citenamefont {Cederstr\"om},\ and\ \citenamefont
  {Imai}}]{HuntPRB2001}%
  \BibitemOpen
  \bibfield  {author} {\bibinfo {author} {\bibfnamefont {A.~W.}\ \bibnamefont
  {Hunt}}, \bibinfo {author} {\bibfnamefont {P.~M.}\ \bibnamefont {Singer}},
  \bibinfo {author} {\bibfnamefont {A.~F.}\ \bibnamefont {Cederstr\"om}},\ and\
  \bibinfo {author} {\bibfnamefont {T.}~\bibnamefont {Imai}},\ }\bibfield
  {title} {\bibinfo {title} {{Glassy slowing of stripe modulation in
  ${(\mathrm{L}\mathrm{a},\mathrm{E}\mathrm{u},\mathrm{N}\mathrm{d})}_{2\ensuremath{-}x}({\mathrm{S}\mathrm{r},\mathrm{B}\mathrm{a})}_{x}{\mathrm{CuO}}_{4}$:
  a ${}^{63}\mathrm{Cu}$ and ${}^{139}\mathrm{La}$ NQR study down to 350 mK}},\
  }\href {https://doi.org/10.1103/PhysRevB.64.134525} {\bibfield  {journal}
  {\bibinfo  {journal} {Phys. Rev. B}\ }\textbf {\bibinfo {volume} {64}},\
  \bibinfo {pages} {134525} (\bibinfo {year} {2001})}\BibitemShut {NoStop}%
\bibitem [{\citenamefont {Fujita}\ \emph {et~al.}(2004)\citenamefont {Fujita},
  \citenamefont {Goka}, \citenamefont {Yamada}, \citenamefont {Tranquada},\
  and\ \citenamefont {Regnault}}]{Fujita2004}%
  \BibitemOpen
  \bibfield  {author} {\bibinfo {author} {\bibfnamefont {M.}~\bibnamefont
  {Fujita}}, \bibinfo {author} {\bibfnamefont {H.}~\bibnamefont {Goka}},
  \bibinfo {author} {\bibfnamefont {K.}~\bibnamefont {Yamada}}, \bibinfo
  {author} {\bibfnamefont {J.~M.}\ \bibnamefont {Tranquada}},\ and\ \bibinfo
  {author} {\bibfnamefont {L.~P.}\ \bibnamefont {Regnault}},\ }\bibfield
  {title} {\bibinfo {title} {{Stripe order, depinning, and fluctuations in
  ${\mathrm{La}}_{1.875}{\mathrm{Ba}}_{0.125}{\mathrm{CuO}}_{4}$ and
  ${\mathrm{La}}_{1.875}{\mathrm{Ba}}_{0.075}{\mathrm{Sr}}_{0.050}{\mathrm{CuO}}_{4}$}},\
  }\href {https://doi.org/10.1103/PhysRevB.70.104517} {\bibfield  {journal}
  {\bibinfo  {journal} {Phys. Rev. B}\ }\textbf {\bibinfo {volume} {70}},\
  \bibinfo {pages} {104517} (\bibinfo {year} {2004})}\BibitemShut {NoStop}%
\bibitem [{\citenamefont {Thampy}\ \emph {et~al.}(2017)\citenamefont {Thampy},
  \citenamefont {Chen}, \citenamefont {Cao}, \citenamefont {Mazzoli},
  \citenamefont {Barbour}, \citenamefont {Hu}, \citenamefont {Miao},
  \citenamefont {Fabbris}, \citenamefont {Zhong}, \citenamefont {Gu},
  \citenamefont {Tranquada}, \citenamefont {Robinson}, \citenamefont
  {Wilkins},\ and\ \citenamefont {Dean}}]{Thampy2017}%
  \BibitemOpen
  \bibfield  {author} {\bibinfo {author} {\bibfnamefont {V.}~\bibnamefont
  {Thampy}}, \bibinfo {author} {\bibfnamefont {X.~M.}\ \bibnamefont {Chen}},
  \bibinfo {author} {\bibfnamefont {Y.}~\bibnamefont {Cao}}, \bibinfo {author}
  {\bibfnamefont {C.}~\bibnamefont {Mazzoli}}, \bibinfo {author} {\bibfnamefont
  {A.~M.}\ \bibnamefont {Barbour}}, \bibinfo {author} {\bibfnamefont
  {W.}~\bibnamefont {Hu}}, \bibinfo {author} {\bibfnamefont {H.}~\bibnamefont
  {Miao}}, \bibinfo {author} {\bibfnamefont {G.}~\bibnamefont {Fabbris}},
  \bibinfo {author} {\bibfnamefont {R.~D.}\ \bibnamefont {Zhong}}, \bibinfo
  {author} {\bibfnamefont {G.~D.}\ \bibnamefont {Gu}}, \bibinfo {author}
  {\bibfnamefont {J.~M.}\ \bibnamefont {Tranquada}}, \bibinfo {author}
  {\bibfnamefont {I.~K.}\ \bibnamefont {Robinson}}, \bibinfo {author}
  {\bibfnamefont {S.~B.}\ \bibnamefont {Wilkins}},\ and\ \bibinfo {author}
  {\bibfnamefont {M.~P.~M.}\ \bibnamefont {Dean}},\ }\bibfield  {title}
  {\bibinfo {title} {{Static charge-density-wave order in the superconducting
  state of
  ${\mathrm{La}}_{2\ensuremath{-}x}{\mathrm{Ba}}_{x}{\mathrm{CuO}}_{4}$}},\
  }\href {https://doi.org/10.1103/PhysRevB.95.241111} {\bibfield  {journal}
  {\bibinfo  {journal} {Phys. Rev. B}\ }\textbf {\bibinfo {volume} {95}},\
  \bibinfo {pages} {241111} (\bibinfo {year} {2017})}\BibitemShut {NoStop}%
\bibitem [{\citenamefont {Miao}\ \emph {et~al.}(2019)\citenamefont {Miao},
  \citenamefont {Fumagalli}, \citenamefont {Rossi}, \citenamefont {Lorenzana},
  \citenamefont {Seibold}, \citenamefont {Yakhou-Harris}, \citenamefont
  {Kummer}, \citenamefont {Brookes}, \citenamefont {Gu}, \citenamefont
  {Braicovich}, \citenamefont {Ghiringhelli},\ and\ \citenamefont
  {Dean}}]{MiaoPRX2019}%
  \BibitemOpen
  \bibfield  {author} {\bibinfo {author} {\bibfnamefont {H.}~\bibnamefont
  {Miao}}, \bibinfo {author} {\bibfnamefont {R.}~\bibnamefont {Fumagalli}},
  \bibinfo {author} {\bibfnamefont {M.}~\bibnamefont {Rossi}}, \bibinfo
  {author} {\bibfnamefont {J.}~\bibnamefont {Lorenzana}}, \bibinfo {author}
  {\bibfnamefont {G.}~\bibnamefont {Seibold}}, \bibinfo {author} {\bibfnamefont
  {F.}~\bibnamefont {Yakhou-Harris}}, \bibinfo {author} {\bibfnamefont
  {K.}~\bibnamefont {Kummer}}, \bibinfo {author} {\bibfnamefont {N.~B.}\
  \bibnamefont {Brookes}}, \bibinfo {author} {\bibfnamefont {G.~D.}\
  \bibnamefont {Gu}}, \bibinfo {author} {\bibfnamefont {L.}~\bibnamefont
  {Braicovich}}, \bibinfo {author} {\bibfnamefont {G.}~\bibnamefont
  {Ghiringhelli}},\ and\ \bibinfo {author} {\bibfnamefont {M.~P.~M.}\
  \bibnamefont {Dean}},\ }\bibfield  {title} {\bibinfo {title} {Formation of
  incommensurate charge density waves in cuprates},\ }\href
  {https://doi.org/10.1103/PhysRevX.9.031042} {\bibfield  {journal} {\bibinfo
  {journal} {Phys. Rev. X}\ }\textbf {\bibinfo {volume} {9}},\ \bibinfo {pages}
  {031042} (\bibinfo {year} {2019})}\BibitemShut {NoStop}%
\bibitem [{\citenamefont {Imai}\ \emph {et~al.}(1990)\citenamefont {Imai},
  \citenamefont {Yoshimura}, \citenamefont {Uemura}, \citenamefont {Yasuoka},\
  and\ \citenamefont {Kosuge}}]{ImaiJPSJ1990}%
  \BibitemOpen
  \bibfield  {author} {\bibinfo {author} {\bibfnamefont {T.}~\bibnamefont
  {Imai}}, \bibinfo {author} {\bibfnamefont {K.}~\bibnamefont {Yoshimura}},
  \bibinfo {author} {\bibfnamefont {T.}~\bibnamefont {Uemura}}, \bibinfo
  {author} {\bibfnamefont {H.}~\bibnamefont {Yasuoka}},\ and\ \bibinfo {author}
  {\bibfnamefont {K.}~\bibnamefont {Kosuge}},\ }\bibfield  {title} {\bibinfo
  {title} {{$^{63}$Cu NMR study of spin dynamics in
  $\mathrm{La}_{2-x}\mathrm{(Sr,Ba)}_{x}\mathrm{CuO}_{y}$ ($0.04 \leq x \leq
  0.16$, $3.99 \leq y \leq 4.03$)}},\ }\href@noop {} {\bibfield  {journal}
  {\bibinfo  {journal} {J. Phys. Soc. Jpn.}\ }\textbf {\bibinfo {volume}
  {59}},\ \bibinfo {pages} {3846} (\bibinfo {year} {1990})}\BibitemShut
  {NoStop}%
\bibitem [{\citenamefont {Tou}\ \emph {et~al.}(1992)\citenamefont {Tou},
  \citenamefont {Matsumura},\ and\ \citenamefont {Yamagata}}]{TouLBCOT2}%
  \BibitemOpen
  \bibfield  {author} {\bibinfo {author} {\bibfnamefont {H.}~\bibnamefont
  {Tou}}, \bibinfo {author} {\bibfnamefont {M.}~\bibnamefont {Matsumura}},\
  and\ \bibinfo {author} {\bibfnamefont {H.}~\bibnamefont {Yamagata}},\
  }\bibfield  {title} {\bibinfo {title} {Anomalous {Cu-NQR} spectral change due
  to low-temperature structural transition around $x=0.06$ in
  {$\mathrm{(La_{1-x}Ba_{x})_{2}CuO_{4}}$}},\ }\href@noop {} {\bibfield
  {journal} {\bibinfo  {journal} {J. Phys. Soc. Jpn.}\ }\textbf {\bibinfo
  {volume} {61}},\ \bibinfo {pages} {1477} (\bibinfo {year}
  {1992})}\BibitemShut {NoStop}%
\bibitem [{\citenamefont {Tou}\ \emph {et~al.}(1993)\citenamefont {Tou},
  \citenamefont {Matsumura},\ and\ \citenamefont {Yamagata}}]{TouLBCOT1}%
  \BibitemOpen
  \bibfield  {author} {\bibinfo {author} {\bibfnamefont {H.}~\bibnamefont
  {Tou}}, \bibinfo {author} {\bibfnamefont {M.}~\bibnamefont {Matsumura}},\
  and\ \bibinfo {author} {\bibfnamefont {H.}~\bibnamefont {Yamagata}},\
  }\bibfield  {title} {\bibinfo {title} {{$\mathrm{^{63}}$Cu} nuclear
  spin-lattice relaxation study for low-temperature structural transition in
  {$\mathrm{(La_{1-x}Ba_{x})_{2}CuO_{4}}$} around $x=0.06$},\ }\href@noop {}
  {\bibfield  {journal} {\bibinfo  {journal} {J. Phys. Soc. Jpn.}\ }\textbf
  {\bibinfo {volume} {62}},\ \bibinfo {pages} {1474} (\bibinfo {year}
  {1993})}\BibitemShut {NoStop}%
\bibitem [{\citenamefont {Kumagai}\ \emph {et~al.}(1994)\citenamefont
  {Kumagai}, \citenamefont {Kawano}, \citenamefont {Watanabe}, \citenamefont
  {Nishiyama},\ and\ \citenamefont {Nagamine}}]{Kumagai1994}%
  \BibitemOpen
  \bibfield  {author} {\bibinfo {author} {\bibfnamefont {K.}~\bibnamefont
  {Kumagai}}, \bibinfo {author} {\bibfnamefont {K.}~\bibnamefont {Kawano}},
  \bibinfo {author} {\bibfnamefont {I.}~\bibnamefont {Watanabe}}, \bibinfo
  {author} {\bibfnamefont {K.}~\bibnamefont {Nishiyama}},\ and\ \bibinfo
  {author} {\bibfnamefont {K.}~\bibnamefont {Nagamine}},\ }\bibfield  {title}
  {\bibinfo {title} {{$\mu$SR and NMR investigations on electronic and magnetic
  state around $x=0.12$ in La$_{2-x}$Sr$_{x}$CuO$_{4}$ and
  La$_{2-x}$Ba$_{x}$CuO$_{4}$}},\ }\href@noop {} {\bibfield  {journal}
  {\bibinfo  {journal} {Hyperfine Interactions}\ }\textbf {\bibinfo {volume}
  {86}},\ \bibinfo {pages} {473} (\bibinfo {year} {1994})}\BibitemShut
  {NoStop}%
\bibitem [{\citenamefont {Goto}\ \emph {et~al.}(1994)\citenamefont {Goto},
  \citenamefont {Kazama}, \citenamefont {Miyagawa},\ and\ \citenamefont
  {Fukase}}]{GotoJPSJ1994}%
  \BibitemOpen
  \bibfield  {author} {\bibinfo {author} {\bibfnamefont {T.}~\bibnamefont
  {Goto}}, \bibinfo {author} {\bibfnamefont {S.}~\bibnamefont {Kazama}},
  \bibinfo {author} {\bibfnamefont {K.}~\bibnamefont {Miyagawa}},\ and\
  \bibinfo {author} {\bibfnamefont {T.}~\bibnamefont {Fukase}},\ }\bibfield
  {title} {\bibinfo {title} {{$\mathrm{^{63/65}Cu/^{139}La}$-NMR study on
  antiferromagnetic ordering in high-$\mathrm{T_{c}}$ oxides
  $\mathrm{La_{2-x}Sr_{x}CuO_{4}}$ ($x \simeq 0.115$) and
  $\mathrm{La_{2-x}Ba_{x}CuO_{4}}$ ($x \simeq 0.125$)}},\ }\href@noop {}
  {\bibfield  {journal} {\bibinfo  {journal} {J. Phys. Soc. Jpn.}\ }\textbf
  {\bibinfo {volume} {63}},\ \bibinfo {pages} {3494} (\bibinfo {year}
  {1994})}\BibitemShut {NoStop}%
\bibitem [{\citenamefont {Baek}\ \emph {et~al.}(2015)\citenamefont {Baek},
  \citenamefont {Utz}, \citenamefont {H\"ucker}, \citenamefont {Gu},
  \citenamefont {B\"uchner},\ and\ \citenamefont {Grafe}}]{BaekLaT1PRB2015}%
  \BibitemOpen
  \bibfield  {author} {\bibinfo {author} {\bibfnamefont {S.-H.}\ \bibnamefont
  {Baek}}, \bibinfo {author} {\bibfnamefont {Y.}~\bibnamefont {Utz}}, \bibinfo
  {author} {\bibfnamefont {M.}~\bibnamefont {H\"ucker}}, \bibinfo {author}
  {\bibfnamefont {G.~D.}\ \bibnamefont {Gu}}, \bibinfo {author} {\bibfnamefont
  {B.}~\bibnamefont {B\"uchner}},\ and\ \bibinfo {author} {\bibfnamefont
  {H.-J.}\ \bibnamefont {Grafe}},\ }\bibfield  {title} {\bibinfo {title}
  {{Magnetic field induced anisotropy of $^{139}\mathrm{La}$ spin-lattice
  relaxation rates in stripe ordered
  ${\mathrm{La}}_{1.875}{\mathrm{Ba}}_{0.125}{\mathrm{CuO}}_{4}$}},\ }\href
  {https://doi.org/10.1103/PhysRevB.92.155144} {\bibfield  {journal} {\bibinfo
  {journal} {Phys. Rev. B}\ }\textbf {\bibinfo {volume} {92}},\ \bibinfo
  {pages} {155144} (\bibinfo {year} {2015})}\BibitemShut {NoStop}%
\bibitem [{\citenamefont {Pelc}\ \emph {et~al.}(2017)\citenamefont {Pelc},
  \citenamefont {Grafe}, \citenamefont {Gu},\ and\ \citenamefont
  {Po\ifmmode~\check{z}\else \v{z}\fi{}ek}}]{Pelc2017}%
  \BibitemOpen
  \bibfield  {author} {\bibinfo {author} {\bibfnamefont {D.}~\bibnamefont
  {Pelc}}, \bibinfo {author} {\bibfnamefont {H.-J.}\ \bibnamefont {Grafe}},
  \bibinfo {author} {\bibfnamefont {G.~D.}\ \bibnamefont {Gu}},\ and\ \bibinfo
  {author} {\bibfnamefont {M.}~\bibnamefont {Po\ifmmode~\check{z}\else
  \v{z}\fi{}ek}},\ }\bibfield  {title} {\bibinfo {title} {{Cu nuclear magnetic
  resonance study of charge and spin stripe order in
  ${\mathrm{La}}_{1.875}{\mathrm{Ba}}_{0.125}{\mathrm{CuO}}_{4}$}},\ }\href
  {https://doi.org/10.1103/PhysRevB.95.054508} {\bibfield  {journal} {\bibinfo
  {journal} {Phys. Rev. B}\ }\textbf {\bibinfo {volume} {95}},\ \bibinfo
  {pages} {054508} (\bibinfo {year} {2017})}\BibitemShut {NoStop}%
\bibitem [{\citenamefont {Johnston}\ \emph {et~al.}(2005)\citenamefont
  {Johnston}, \citenamefont {Baek}, \citenamefont {Zong}, \citenamefont
  {Borsa}, \citenamefont {Schmalian},\ and\ \citenamefont
  {Kondo}}]{JohnstonPRL2005}%
  \BibitemOpen
  \bibfield  {author} {\bibinfo {author} {\bibfnamefont {D.~C.}\ \bibnamefont
  {Johnston}}, \bibinfo {author} {\bibfnamefont {S.-H.}\ \bibnamefont {Baek}},
  \bibinfo {author} {\bibfnamefont {X.}~\bibnamefont {Zong}}, \bibinfo {author}
  {\bibfnamefont {F.}~\bibnamefont {Borsa}}, \bibinfo {author} {\bibfnamefont
  {J.}~\bibnamefont {Schmalian}},\ and\ \bibinfo {author} {\bibfnamefont
  {S.}~\bibnamefont {Kondo}},\ }\bibfield  {title} {\bibinfo {title} {{Dynamics
  of magnetic defects in heavy fermion ${\mathrm{LiV}}_{2}{\mathrm{O}}_{4}$
  from stretched exponential $^{7}\mathrm{Li}$ NMR relaxation}},\ }\href
  {https://doi.org/10.1103/PhysRevLett.95.176408} {\bibfield  {journal}
  {\bibinfo  {journal} {Phys. Rev. Lett.}\ }\textbf {\bibinfo {volume} {95}},\
  \bibinfo {pages} {176408} (\bibinfo {year} {2005})}\BibitemShut {NoStop}%
\bibitem [{\citenamefont {Venkataramanan}\ \emph {et~al.}(2002)\citenamefont
  {Venkataramanan}, \citenamefont {Song},\ and\ \citenamefont
  {H{\"u}rlimann}}]{venkataramanan:ieee2002}%
  \BibitemOpen
  \bibfield  {author} {\bibinfo {author} {\bibfnamefont {L.}~\bibnamefont
  {Venkataramanan}}, \bibinfo {author} {\bibfnamefont {Y.-Q.}\ \bibnamefont
  {Song}},\ and\ \bibinfo {author} {\bibfnamefont {M.~D.}\ \bibnamefont
  {H{\"u}rlimann}},\ }\bibfield  {title} {\bibinfo {title} {Solving fredholm
  integrals of the first kind with tensor product structure in 2 and 2.5
  dimensions},\ }\href@noop {} {\bibfield  {journal} {\bibinfo  {journal} {IEEE
  Trans. Sig. Process.}\ }\textbf {\bibinfo {volume} {50 (5)}},\ \bibinfo
  {pages} {1017} (\bibinfo {year} {2002})}\BibitemShut {NoStop}%
\bibitem [{\citenamefont {Song}\ \emph {et~al.}(2002)\citenamefont {Song},
  \citenamefont {Venkataramanan}, \citenamefont {H{\"u}rlimann}, \citenamefont
  {Flaum}, \citenamefont {Frulla},\ and\ \citenamefont
  {Straley}}]{song:jmr2002}%
  \BibitemOpen
  \bibfield  {author} {\bibinfo {author} {\bibfnamefont {Y.-Q.}\ \bibnamefont
  {Song}}, \bibinfo {author} {\bibfnamefont {L.}~\bibnamefont
  {Venkataramanan}}, \bibinfo {author} {\bibfnamefont {M.~D.}\ \bibnamefont
  {H{\"u}rlimann}}, \bibinfo {author} {\bibfnamefont {M.}~\bibnamefont
  {Flaum}}, \bibinfo {author} {\bibfnamefont {P.}~\bibnamefont {Frulla}},\ and\
  \bibinfo {author} {\bibfnamefont {C.}~\bibnamefont {Straley}},\ }\bibfield
  {title} {\bibinfo {title} {{$T_1$-$T_2$} correlation spectra obtained using
  fast two-dimensional laplace inversion},\ }\href@noop {} {\bibfield
  {journal} {\bibinfo  {journal} {J. Magn. Reson.}\ }\textbf {\bibinfo {volume}
  {154}},\ \bibinfo {pages} {261} (\bibinfo {year} {2002})}\BibitemShut
  {NoStop}%
\bibitem [{\citenamefont {Mitchell}\ \emph {et~al.}(2012)\citenamefont
  {Mitchell}, \citenamefont {Chandrasekera},\ and\ \citenamefont
  {Gladden}}]{mitchell:PNMRS2012}%
  \BibitemOpen
  \bibfield  {author} {\bibinfo {author} {\bibfnamefont {J.}~\bibnamefont
  {Mitchell}}, \bibinfo {author} {\bibfnamefont {T.~C.}\ \bibnamefont
  {Chandrasekera}},\ and\ \bibinfo {author} {\bibfnamefont {L.}~\bibnamefont
  {Gladden}},\ }\bibfield  {title} {\bibinfo {title} {Numerical estimation of
  relaxation and diffusion distributions in two dimensions},\ }\href@noop {}
  {\bibfield  {journal} {\bibinfo  {journal} {Prog. Nucl. Magn. Reson. Spect.}\
  }\textbf {\bibinfo {volume} {62}},\ \bibinfo {pages} {34} (\bibinfo {year}
  {2012})}\BibitemShut {NoStop}%
\bibitem [{\citenamefont {Singer}\ \emph
  {et~al.}(2018{\natexlab{a}})\citenamefont {Singer}, \citenamefont
  {Asthagiri}, \citenamefont {Chen}, \citenamefont {{Valiya Parambathu}},
  \citenamefont {Hirasaki},\ and\ \citenamefont {Chapman}}]{singer:jcp2018}%
  \BibitemOpen
  \bibfield  {author} {\bibinfo {author} {\bibfnamefont {P.~M.}\ \bibnamefont
  {Singer}}, \bibinfo {author} {\bibfnamefont {D.}~\bibnamefont {Asthagiri}},
  \bibinfo {author} {\bibfnamefont {Z.}~\bibnamefont {Chen}}, \bibinfo {author}
  {\bibfnamefont {A.}~\bibnamefont {{Valiya Parambathu}}}, \bibinfo {author}
  {\bibfnamefont {G.~J.}\ \bibnamefont {Hirasaki}},\ and\ \bibinfo {author}
  {\bibfnamefont {W.~G.}\ \bibnamefont {Chapman}},\ }\bibfield  {title}
  {\bibinfo {title} {Role of internal motions and molecular geometry on the
  {NMR} relaxation of hydrocarbons},\ }\href@noop {} {\bibfield  {journal}
  {\bibinfo  {journal} {J. Chem. Phys.}\ }\textbf {\bibinfo {volume} {148
  (16)}},\ \bibinfo {pages} {164507} (\bibinfo {year}
  {2018}{\natexlab{a}})}\BibitemShut {NoStop}%
\bibitem [{\citenamefont {Singer}\ \emph
  {et~al.}(2018{\natexlab{b}})\citenamefont {Singer}, \citenamefont
  {Asthagiri}, \citenamefont {Chapman},\ and\ \citenamefont
  {Hirasaki}}]{singer:jcp2018b}%
  \BibitemOpen
  \bibfield  {author} {\bibinfo {author} {\bibfnamefont {P.~M.}\ \bibnamefont
  {Singer}}, \bibinfo {author} {\bibfnamefont {D.}~\bibnamefont {Asthagiri}},
  \bibinfo {author} {\bibfnamefont {W.~G.}\ \bibnamefont {Chapman}},\ and\
  \bibinfo {author} {\bibfnamefont {G.~J.}\ \bibnamefont {Hirasaki}},\
  }\bibfield  {title} {\bibinfo {title} {{NMR} spin-rotation relaxation and
  diffusion of methane},\ }\href@noop {} {\bibfield  {journal} {\bibinfo
  {journal} {J. Chem. Phys.}\ }\textbf {\bibinfo {volume} {148 (20)}},\
  \bibinfo {pages} {204504} (\bibinfo {year} {2018}{\natexlab{b}})}\BibitemShut
  {NoStop}%
\bibitem [{\citenamefont {Arsenault}\ \emph {et~al.}()\citenamefont
  {Arsenault}, \citenamefont {Imai}, \citenamefont {Singer}, \citenamefont
  {Suzuki},\ and\ \citenamefont {Fujita}}]{ArsenaultPRB2019}%
  \BibitemOpen
  \bibfield  {author} {\bibinfo {author} {\bibfnamefont {A.}~\bibnamefont
  {Arsenault}}, \bibinfo {author} {\bibfnamefont {T.}~\bibnamefont {Imai}},
  \bibinfo {author} {\bibfnamefont {P.~M.}\ \bibnamefont {Singer}}, \bibinfo
  {author} {\bibfnamefont {K.~M.}\ \bibnamefont {Suzuki}},\ and\ \bibinfo
  {author} {\bibfnamefont {M.}~\bibnamefont {Fujita}},\ }\bibfield  {title}
  {\bibinfo {title} {{Magnetic inhomogeneity in charge ordered
  La$_{1.885}$Sr$_{0.115}$CuO$_4$} investigated by {NMR}},\ }\bibinfo {note}
  {accepted for publication in Phys. Bev. B (arXiv:1912.11448)}\BibitemShut
  {NoStop}%
\bibitem [{\citenamefont {Takahashi}\ \emph {et~al.}(2019)\citenamefont
  {Takahashi}, \citenamefont {Wang}, \citenamefont {Arsenault}, \citenamefont
  {Imai}, \citenamefont {Abramchuk}, \citenamefont {Tafti},\ and\ \citenamefont
  {Singer}}]{TakahashiPRX2019}%
  \BibitemOpen
  \bibfield  {author} {\bibinfo {author} {\bibfnamefont {S.~K.}\ \bibnamefont
  {Takahashi}}, \bibinfo {author} {\bibfnamefont {J.}~\bibnamefont {Wang}},
  \bibinfo {author} {\bibfnamefont {A.}~\bibnamefont {Arsenault}}, \bibinfo
  {author} {\bibfnamefont {T.}~\bibnamefont {Imai}}, \bibinfo {author}
  {\bibfnamefont {M.}~\bibnamefont {Abramchuk}}, \bibinfo {author}
  {\bibfnamefont {F.}~\bibnamefont {Tafti}},\ and\ \bibinfo {author}
  {\bibfnamefont {P.~M.}\ \bibnamefont {Singer}},\ }\bibfield  {title}
  {\bibinfo {title} {Spin excitations of a proximate kitaev quantum spin liquid
  realized in {${\mathrm{Cu}}_{2}{\mathrm{IrO}}_{3}$}},\ }\href
  {https://doi.org/10.1103/PhysRevX.9.031047} {\bibfield  {journal} {\bibinfo
  {journal} {Phys. Rev. X}\ }\textbf {\bibinfo {volume} {9}},\ \bibinfo {pages}
  {031047} (\bibinfo {year} {2019})}\BibitemShut {NoStop}%
\bibitem [{\citenamefont {Itoh}\ \emph {et~al.}(1986)\citenamefont {Itoh},
  \citenamefont {Yasuoka}, \citenamefont {King},\ and\ \citenamefont
  {Jaccarino}}]{Itoh1986}%
  \BibitemOpen
  \bibfield  {author} {\bibinfo {author} {\bibfnamefont {M.}~\bibnamefont
  {Itoh}}, \bibinfo {author} {\bibfnamefont {H.}~\bibnamefont {Yasuoka}},
  \bibinfo {author} {\bibfnamefont {A.~R.}\ \bibnamefont {King}},\ and\
  \bibinfo {author} {\bibfnamefont {V.}~\bibnamefont {Jaccarino}},\ }\bibfield
  {title} {\bibinfo {title} {{Decay of the Nuclear Magnetization in the
  Randomly Diluted Antiferromagnets Fe$_x$Zn$_{1-x}$F$_2$ and
  Mn$_x$Zn$_{1-x}$F$_2$}},\ }\href@noop {} {\bibfield  {journal} {\bibinfo
  {journal} {J. Phys. Soc. Jpn.}\ }\textbf {\bibinfo {volume} {55}},\ \bibinfo
  {pages} {964} (\bibinfo {year} {1986})}\BibitemShut {NoStop}%
\bibitem [{\citenamefont {Thayamballi}\ and\ \citenamefont
  {Hone}(1980)}]{Thayamballi1980}%
  \BibitemOpen
  \bibfield  {author} {\bibinfo {author} {\bibfnamefont {P.}~\bibnamefont
  {Thayamballi}}\ and\ \bibinfo {author} {\bibfnamefont {D.}~\bibnamefont
  {Hone}},\ }\bibfield  {title} {\bibinfo {title} {{Nuclear Relaxation in a
  Randomly Diluted Heisenberg Paramagnet}},\ }\href
  {https://doi.org/10.1103/PhysRevB.21.1766} {\bibfield  {journal} {\bibinfo
  {journal} {Phys. Rev. B}\ }\textbf {\bibinfo {volume} {21}},\ \bibinfo
  {pages} {1766} (\bibinfo {year} {1980})}\BibitemShut {NoStop}%
\bibitem [{\citenamefont {Imai}\ \emph
  {et~al.}(1988{\natexlab{a}})\citenamefont {Imai}, \citenamefont {Shimizu},
  \citenamefont {Yasuoka}, \citenamefont {Ueda},\ and\ \citenamefont
  {Kosuge}}]{ImaiJPSJ1988}%
  \BibitemOpen
  \bibfield  {author} {\bibinfo {author} {\bibfnamefont {T.}~\bibnamefont
  {Imai}}, \bibinfo {author} {\bibfnamefont {T.}~\bibnamefont {Shimizu}},
  \bibinfo {author} {\bibfnamefont {H.}~\bibnamefont {Yasuoka}}, \bibinfo
  {author} {\bibfnamefont {Y.}~\bibnamefont {Ueda}},\ and\ \bibinfo {author}
  {\bibfnamefont {K.}~\bibnamefont {Kosuge}},\ }\bibfield  {title} {\bibinfo
  {title} {{Nuclear spin-lattice relaxation of $^{63,65}\mathrm{Cu}$ at the
  Cu(2) sites of the high $T_{c}$ superconductor
  $\mathrm{YBa}_{2}\mathrm{Cu}_{3}\mathrm{O}_{7-\delta}$}},\ }\href@noop {}
  {\bibfield  {journal} {\bibinfo  {journal} {J. Phys. Soc. Jpn.}\ }\textbf
  {\bibinfo {volume} {57}},\ \bibinfo {pages} {2280} (\bibinfo {year}
  {1988}{\natexlab{a}})}\BibitemShut {NoStop}%
\bibitem [{\citenamefont {Imai}\ \emph
  {et~al.}(1988{\natexlab{b}})\citenamefont {Imai}, \citenamefont {Shimizu},
  \citenamefont {Yasuoka}, \citenamefont {Ueda},\ and\ \citenamefont
  {Kosuge}}]{ImaiJPSJ1988-2}%
  \BibitemOpen
  \bibfield  {author} {\bibinfo {author} {\bibfnamefont {T.}~\bibnamefont
  {Imai}}, \bibinfo {author} {\bibfnamefont {T.}~\bibnamefont {Shimizu}},
  \bibinfo {author} {\bibfnamefont {H.}~\bibnamefont {Yasuoka}}, \bibinfo
  {author} {\bibfnamefont {Y.}~\bibnamefont {Ueda}},\ and\ \bibinfo {author}
  {\bibfnamefont {K.}~\bibnamefont {Kosuge}},\ }\bibfield  {title} {\bibinfo
  {title} {{Anomalous Temperature dependence of Cu Nuclear Spin-Lattice
  Relaxation in $\mathrm{YBa}_{2}\mathrm{Cu}_{3}\mathrm{O}_{6.91}$}},\
  }\href@noop {} {\bibfield  {journal} {\bibinfo  {journal} {J. Phys. Soc.
  Jpn.}\ }\textbf {\bibinfo {volume} {57}},\ \bibinfo {pages} {2280} (\bibinfo
  {year} {1988}{\natexlab{b}})}\BibitemShut {NoStop}%
\bibitem [{\citenamefont {Imai}\ \emph {et~al.}(1989)\citenamefont {Imai},
  \citenamefont {Yasuoka}, \citenamefont {Shimizu}, \citenamefont {Ueda},
  \citenamefont {Yoshimura},\ and\ \citenamefont {Kosuge}}]{ImaiPhysicaC1989}%
  \BibitemOpen
  \bibfield  {author} {\bibinfo {author} {\bibfnamefont {T.}~\bibnamefont
  {Imai}}, \bibinfo {author} {\bibfnamefont {H.}~\bibnamefont {Yasuoka}},
  \bibinfo {author} {\bibfnamefont {T.}~\bibnamefont {Shimizu}}, \bibinfo
  {author} {\bibfnamefont {Y.}~\bibnamefont {Ueda}}, \bibinfo {author}
  {\bibfnamefont {K.}~\bibnamefont {Yoshimura}},\ and\ \bibinfo {author}
  {\bibfnamefont {K.}~\bibnamefont {Kosuge}},\ }\bibfield  {title} {\bibinfo
  {title} {{Cu spin dyanmics in high $T_c$ and related oxides investigated by
  nuclear spin-lattice relaxation}},\ }\href@noop {} {\bibfield  {journal}
  {\bibinfo  {journal} {Physica C}\ }\textbf {\bibinfo {volume} {162-164}},\
  \bibinfo {pages} {169} (\bibinfo {year} {1989})}\BibitemShut {NoStop}%
\bibitem [{\citenamefont {Andrew}\ and\ \citenamefont
  {Tunstall}(1961)}]{Andrew1961}%
  \BibitemOpen
  \bibfield  {author} {\bibinfo {author} {\bibfnamefont {E.~R.}\ \bibnamefont
  {Andrew}}\ and\ \bibinfo {author} {\bibfnamefont {D.~P.}\ \bibnamefont
  {Tunstall}},\ }\bibfield  {title} {\bibinfo {title} {{Spin-Lattice Relaxation
  in Imperfect Cubic Crystals and in Non-cubic Crystals}},\ }\href@noop {}
  {\bibfield  {journal} {\bibinfo  {journal} {Proceeings of the Royal Society}\
  }\textbf {\bibinfo {volume} {78}},\ \bibinfo {pages} {1} (\bibinfo {year}
  {1961})}\BibitemShut {NoStop}%
\bibitem [{\citenamefont {Narath}(1967)}]{Narath1967}%
  \BibitemOpen
  \bibfield  {author} {\bibinfo {author} {\bibfnamefont {A.}~\bibnamefont
  {Narath}},\ }\bibfield  {title} {\bibinfo {title} {Nuclear spin-lattice
  relaxation in hexagonal transition metals: Titanium},\ }\href
  {https://doi.org/10.1103/PhysRev.162.320} {\bibfield  {journal} {\bibinfo
  {journal} {Phys. Rev.}\ }\textbf {\bibinfo {volume} {162}},\ \bibinfo {pages}
  {320} (\bibinfo {year} {1967})}\BibitemShut {NoStop}%
\bibitem [{\citenamefont {Singer}\ \emph
  {et~al.}(2018{\natexlab{c}})\citenamefont {Singer}, \citenamefont
  {Asthagiri}, \citenamefont {Chapman},\ and\ \citenamefont
  {Hirasaki}}]{SingerJCP2018}%
  \BibitemOpen
  \bibfield  {author} {\bibinfo {author} {\bibfnamefont {P.~M.}\ \bibnamefont
  {Singer}}, \bibinfo {author} {\bibfnamefont {D.}~\bibnamefont {Asthagiri}},
  \bibinfo {author} {\bibfnamefont {W.~G.}\ \bibnamefont {Chapman}},\ and\
  \bibinfo {author} {\bibfnamefont {G.~J.}\ \bibnamefont {Hirasaki}},\
  }\bibfield  {title} {\bibinfo {title} {{NMR} spin-rotation relaxation and
  diffusion of methane},\ }\href {https://doi.org/10.1063/1.5027097} {\bibfield
   {journal} {\bibinfo  {journal} {The Journal of Chemical Physics}\ }\textbf
  {\bibinfo {volume} {148}},\ \bibinfo {pages} {204504} (\bibinfo {year}
  {2018}{\natexlab{c}})},\ \Eprint
  {https://arxiv.org/abs/https://doi.org/10.1063/1.5027097}
  {https://doi.org/10.1063/1.5027097} \BibitemShut {NoStop}%
\bibitem [{Sup()}]{SuppMat}%
  \BibitemOpen
  \href@noop {} {\bibinfo {title} {See supplementary material available at
  [url] for technical details of {ILT}.}}\BibitemShut {Stop}%
\bibitem [{\citenamefont {Chouzenoux}\ \emph {et~al.}(2010)\citenamefont
  {Chouzenoux}, \citenamefont {Moussaoui}, \citenamefont {Idier},\ and\
  \citenamefont {Mariette}}]{chouzenoux:ieee2010}%
  \BibitemOpen
  \bibfield  {author} {\bibinfo {author} {\bibfnamefont {E.}~\bibnamefont
  {Chouzenoux}}, \bibinfo {author} {\bibfnamefont {S.}~\bibnamefont
  {Moussaoui}}, \bibinfo {author} {\bibfnamefont {J.}~\bibnamefont {Idier}},\
  and\ \bibinfo {author} {\bibfnamefont {F.}~\bibnamefont {Mariette}},\
  }\bibfield  {title} {\bibinfo {title} {Efficient maximum entropy
  reconstruction of nuclear magnetic resonance {$T_1$-$T_2$} spectra},\
  }\href@noop {} {\bibfield  {journal} {\bibinfo  {journal} {IEEE Trans. Sig.
  Process.}\ }\textbf {\bibinfo {volume} {58 (12)}},\ \bibinfo {pages} {6040}
  (\bibinfo {year} {2010})}\BibitemShut {NoStop}%
\bibitem [{\citenamefont {Prange}\ and\ \citenamefont
  {Song}(2009)}]{prange:JMR2009}%
  \BibitemOpen
  \bibfield  {author} {\bibinfo {author} {\bibfnamefont {M.}~\bibnamefont
  {Prange}}\ and\ \bibinfo {author} {\bibfnamefont {Y.-Q.}\ \bibnamefont
  {Song}},\ }\bibfield  {title} {\bibinfo {title} {Quantifying uncertainty in
  {NMR} {$T_2$} spectra using {Monte Carlo} inversion},\ }\href@noop {}
  {\bibfield  {journal} {\bibinfo  {journal} {J. Magn. Reson.}\ }\textbf
  {\bibinfo {volume} {196}},\ \bibinfo {pages} {54} (\bibinfo {year}
  {2009})}\BibitemShut {NoStop}%
\bibitem [{\citenamefont {Venkataramanan}\ \emph {et~al.}(2010)\citenamefont
  {Venkataramanan}, \citenamefont {Gruber}, \citenamefont {Habashy},\ and\
  \citenamefont {Freed}}]{venkataramanan:jmr2010}%
  \BibitemOpen
  \bibfield  {author} {\bibinfo {author} {\bibfnamefont {L.}~\bibnamefont
  {Venkataramanan}}, \bibinfo {author} {\bibfnamefont {F.~K.}\ \bibnamefont
  {Gruber}}, \bibinfo {author} {\bibfnamefont {T.~M.}\ \bibnamefont
  {Habashy}},\ and\ \bibinfo {author} {\bibfnamefont {D.~E.}\ \bibnamefont
  {Freed}},\ }\bibfield  {title} {\bibinfo {title} {Mellin transform of {CPMG}
  data},\ }\href@noop {} {\bibfield  {journal} {\bibinfo  {journal} {J. Magn.
  Reson.}\ }\textbf {\bibinfo {volume} {206}},\ \bibinfo {pages} {20} (\bibinfo
  {year} {2010})}\BibitemShut {NoStop}%
\bibitem [{\citenamefont {Fordham}\ \emph {et~al.}(2017)\citenamefont
  {Fordham}, \citenamefont {Venkataramanan}, \citenamefont {Mitchell},\ and\
  \citenamefont {Valori}}]{fordham:diff2017}%
  \BibitemOpen
  \bibfield  {author} {\bibinfo {author} {\bibfnamefont {E.~J.}\ \bibnamefont
  {Fordham}}, \bibinfo {author} {\bibfnamefont {L.}~\bibnamefont
  {Venkataramanan}}, \bibinfo {author} {\bibfnamefont {J.}~\bibnamefont
  {Mitchell}},\ and\ \bibinfo {author} {\bibfnamefont {A.}~\bibnamefont
  {Valori}},\ }\bibfield  {title} {\bibinfo {title} {What are, and what are not
  inverse {Laplace} transforms},\ }\href@noop {} {\bibfield  {journal}
  {\bibinfo  {journal} {Diffusion Fundam.}\ }\textbf {\bibinfo {volume} {29}},\
  \bibinfo {pages} {1} (\bibinfo {year} {2017})}\BibitemShut {NoStop}%
\bibitem [{\citenamefont {Baek}\ \emph {et~al.}(2017)\citenamefont {Baek},
  \citenamefont {Erb},\ and\ \citenamefont {B\"uchner}}]{BaekPRB2017}%
  \BibitemOpen
  \bibfield  {author} {\bibinfo {author} {\bibfnamefont {S.-H.}\ \bibnamefont
  {Baek}}, \bibinfo {author} {\bibfnamefont {A.}~\bibnamefont {Erb}},\ and\
  \bibinfo {author} {\bibfnamefont {B.}~\bibnamefont {B\"uchner}},\ }\bibfield
  {title} {\bibinfo {title} {{Low-energy spin dynamics and critical hole
  concentrations in
  ${\mathrm{La}}_{2\ensuremath{-}x}{\mathrm{Sr}}_{x}{\mathrm{CuO}}_{4}$
  $(0.07\ensuremath{\le}x\ensuremath{\le}0.2)$ revealed by $^{139}\mathrm{La}$
  and $^{63}\mathrm{Cu}$ nuclear magnetic resonance}},\ }\href
  {https://doi.org/10.1103/PhysRevB.96.094519} {\bibfield  {journal} {\bibinfo
  {journal} {Phys. Rev. B}\ }\textbf {\bibinfo {volume} {96}},\ \bibinfo
  {pages} {094519} (\bibinfo {year} {2017})}\BibitemShut {NoStop}%
\bibitem [{\citenamefont {Tranquada}\ \emph {et~al.}(1999)\citenamefont
  {Tranquada}, \citenamefont {Ichikawa},\ and\ \citenamefont
  {Uchida}}]{TranquadaPRB59}%
  \BibitemOpen
  \bibfield  {author} {\bibinfo {author} {\bibfnamefont {J.~M.}\ \bibnamefont
  {Tranquada}}, \bibinfo {author} {\bibfnamefont {N.}~\bibnamefont
  {Ichikawa}},\ and\ \bibinfo {author} {\bibfnamefont {S.}~\bibnamefont
  {Uchida}},\ }\bibfield  {title} {\bibinfo {title} {{Glassy nature of stripe
  ordering in
  ${\mathrm{La}}_{1.6\ensuremath{-}x}{\mathrm{Nd}}_{0.4}{\mathrm{Sr}}_{x}{\mathrm{CuO}}_{4}$}},\
  }\href {https://doi.org/10.1103/PhysRevB.59.14712} {\bibfield  {journal}
  {\bibinfo  {journal} {Phys. Rev. B}\ }\textbf {\bibinfo {volume} {59}},\
  \bibinfo {pages} {14712} (\bibinfo {year} {1999})}\BibitemShut {NoStop}%
\bibitem [{\citenamefont {Imai}\ \emph {et~al.}(2019)\citenamefont {Imai},
  \citenamefont {Arsenault}, \citenamefont {Singer},\ and\ \citenamefont
  {Fujita}}]{ImaiPRB2019}%
  \BibitemOpen
  \bibfield  {author} {\bibinfo {author} {\bibfnamefont {T.}~\bibnamefont
  {Imai}}, \bibinfo {author} {\bibfnamefont {A.}~\bibnamefont {Arsenault}},
  \bibinfo {author} {\bibfnamefont {P.~M.}\ \bibnamefont {Singer}},\ and\
  \bibinfo {author} {\bibfnamefont {M.}~\bibnamefont {Fujita}},\ }\bibfield
  {title} {\bibinfo {title} {{Revisitng $^{63}$Cu NMR evidence for charge order
  in La$_{1.875}$Ba$_{0.125}$CuO$_4$}},\ }\href@noop {} {\bibfield  {journal}
  {\bibinfo  {journal} {Submitted to Phys. Rev. B}\ } (\bibinfo {year}
  {2019})}\BibitemShut {NoStop}%
\bibitem [{\citenamefont {Suter}\ \emph {et~al.}(1998)\citenamefont {Suter},
  \citenamefont {Mali}, \citenamefont {Roos},\ and\ \citenamefont
  {Brinkmann}}]{Suter1998}%
  \BibitemOpen
  \bibfield  {author} {\bibinfo {author} {\bibfnamefont {A.}~\bibnamefont
  {Suter}}, \bibinfo {author} {\bibfnamefont {M.}~\bibnamefont {Mali}},
  \bibinfo {author} {\bibfnamefont {J.}~\bibnamefont {Roos}},\ and\ \bibinfo
  {author} {\bibfnamefont {D.}~\bibnamefont {Brinkmann}},\ }\bibfield  {title}
  {\bibinfo {title} {Mixed magnetic and quadrupolar relaxation in the presence
  of a dominant static zeeman hamiltonian},\ }\href
  {https://doi.org/10.1088/0953-8984/10/26/022} {\bibfield  {journal} {\bibinfo
   {journal} {Journal of Physics: Condensed Matter}\ }\textbf {\bibinfo
  {volume} {10}},\ \bibinfo {pages} {5977} (\bibinfo {year}
  {1998})}\BibitemShut {NoStop}%
\bibitem [{\citenamefont {Kobayashi}\ \emph {et~al.}(1989)\citenamefont
  {Kobayashi}, \citenamefont {Wada}, \citenamefont {Kitaoka},\ and\
  \citenamefont {Asayama}}]{Kobayashi1989}%
  \BibitemOpen
  \bibfield  {author} {\bibinfo {author} {\bibfnamefont {T.}~\bibnamefont
  {Kobayashi}}, \bibinfo {author} {\bibfnamefont {S.}~\bibnamefont {Wada}},
  \bibinfo {author} {\bibfnamefont {Y.}~\bibnamefont {Kitaoka}},\ and\ \bibinfo
  {author} {\bibfnamefont {K.}~\bibnamefont {Asayama}},\ }\bibfield  {title}
  {\bibinfo {title} {{Nuclear spin-lattice relaxation of $^{139}\mathrm{La}$ in
  superconducting $\mathrm{(La_{1-x}Sr_{x})}_{2}\mathrm{CuO}_{4}$}},\
  }\href@noop {} {\bibfield  {journal} {\bibinfo  {journal} {J. Phys. Soc.
  Jpn.}\ }\textbf {\bibinfo {volume} {58}},\ \bibinfo {pages} {2662} (\bibinfo
  {year} {1989})}\BibitemShut {NoStop}%
\bibitem [{\citenamefont {Yoshimura}\ \emph {et~al.}(1992)\citenamefont
  {Yoshimura}, \citenamefont {Uemura}, \citenamefont {Kato}, \citenamefont
  {Shibata}, \citenamefont {Kosuge}, \citenamefont {Imai},\ and\ \citenamefont
  {Yasuoka}}]{Yoshimura1992}%
  \BibitemOpen
  \bibfield  {author} {\bibinfo {author} {\bibfnamefont {K.}~\bibnamefont
  {Yoshimura}}, \bibinfo {author} {\bibfnamefont {T.}~\bibnamefont {Uemura}},
  \bibinfo {author} {\bibfnamefont {M.}~\bibnamefont {Kato}}, \bibinfo {author}
  {\bibfnamefont {T.}~\bibnamefont {Shibata}}, \bibinfo {author} {\bibfnamefont
  {K.}~\bibnamefont {Kosuge}}, \bibinfo {author} {\bibfnamefont
  {T.}~\bibnamefont {Imai}},\ and\ \bibinfo {author} {\bibfnamefont
  {H.}~\bibnamefont {Yasuoka}},\ }\bibfield  {title} {\bibinfo {title}
  {Magnetic phase differentiation in the {La$_{2-x}$Sr$_{x}$CuO$_{y}$} systems
  {-Cu and La} nuclear quadrupole resonance and relaxation},\ }\href@noop {}
  {\bibfield  {journal} {\bibinfo  {journal} {Springer Proceedings in Physics}\
  }\textbf {\bibinfo {volume} {60}},\ \bibinfo {pages} {405} (\bibinfo {year}
  {1992})}\BibitemShut {NoStop}%
\bibitem [{\citenamefont {Singer}\ \emph {et~al.}(2002)\citenamefont {Singer},
  \citenamefont {Hunt},\ and\ \citenamefont {Imai}}]{SingerPRL2002}%
  \BibitemOpen
  \bibfield  {author} {\bibinfo {author} {\bibfnamefont {P.~M.}\ \bibnamefont
  {Singer}}, \bibinfo {author} {\bibfnamefont {A.~W.}\ \bibnamefont {Hunt}},\
  and\ \bibinfo {author} {\bibfnamefont {T.}~\bibnamefont {Imai}},\ }\bibfield
  {title} {\bibinfo {title} {${}^{63}\mathrm{Cu}$ {NQR} evidence for spatial
  variation of hole concentration in
  {${\mathrm{La}}_{2\ensuremath{-}\mathit{x}}{\mathrm{Sr}}_{\mathit{x}}{\mathrm{CuO}}_{4}$}},\
  }\href {https://doi.org/10.1103/PhysRevLett.88.047602} {\bibfield  {journal}
  {\bibinfo  {journal} {Phys. Rev. Lett.}\ }\textbf {\bibinfo {volume} {88}},\
  \bibinfo {pages} {047602} (\bibinfo {year} {2002})}\BibitemShut {NoStop}%
\bibitem [{\citenamefont {Croft}\ \emph {et~al.}(2014)\citenamefont {Croft},
  \citenamefont {Lester}, \citenamefont {Senn}, \citenamefont {Bombardi},\ and\
  \citenamefont {Hayden}}]{Croft}%
  \BibitemOpen
  \bibfield  {author} {\bibinfo {author} {\bibfnamefont {T.~P.}\ \bibnamefont
  {Croft}}, \bibinfo {author} {\bibfnamefont {C.}~\bibnamefont {Lester}},
  \bibinfo {author} {\bibfnamefont {M.~S.}\ \bibnamefont {Senn}}, \bibinfo
  {author} {\bibfnamefont {A.}~\bibnamefont {Bombardi}},\ and\ \bibinfo
  {author} {\bibfnamefont {S.~M.}\ \bibnamefont {Hayden}},\ }\bibfield  {title}
  {\bibinfo {title} {{Charge density wave fluctuations in
  ${\text{La}}_{2\ensuremath{-}x}$${\text{Sr}}_{x}$${\text{CuO}}_{4}$ and their
  competition with superconductivity}},\ }\href
  {https://doi.org/10.1103/PhysRevB.89.224513} {\bibfield  {journal} {\bibinfo
  {journal} {Phys. Rev. B}\ }\textbf {\bibinfo {volume} {89}},\ \bibinfo
  {pages} {224513} (\bibinfo {year} {2014})}\BibitemShut {NoStop}%
\bibitem [{\citenamefont {Thampy}\ \emph {et~al.}(2014)\citenamefont {Thampy},
  \citenamefont {Dean}, \citenamefont {Christensen}, \citenamefont {Steinke},
  \citenamefont {Islam}, \citenamefont {Oda}, \citenamefont {Ido},
  \citenamefont {Momono}, \citenamefont {Wilkins},\ and\ \citenamefont
  {Hill}}]{Thampy}%
  \BibitemOpen
  \bibfield  {author} {\bibinfo {author} {\bibfnamefont {V.}~\bibnamefont
  {Thampy}}, \bibinfo {author} {\bibfnamefont {M.~P.~M.}\ \bibnamefont {Dean}},
  \bibinfo {author} {\bibfnamefont {N.~B.}\ \bibnamefont {Christensen}},
  \bibinfo {author} {\bibfnamefont {L.}~\bibnamefont {Steinke}}, \bibinfo
  {author} {\bibfnamefont {Z.}~\bibnamefont {Islam}}, \bibinfo {author}
  {\bibfnamefont {M.}~\bibnamefont {Oda}}, \bibinfo {author} {\bibfnamefont
  {M.}~\bibnamefont {Ido}}, \bibinfo {author} {\bibfnamefont {N.}~\bibnamefont
  {Momono}}, \bibinfo {author} {\bibfnamefont {S.~B.}\ \bibnamefont
  {Wilkins}},\ and\ \bibinfo {author} {\bibfnamefont {J.~P.}\ \bibnamefont
  {Hill}},\ }\bibfield  {title} {\bibinfo {title} {{Rotated stripe order and
  its competition with superconductivity in
  ${\mathrm{La}}_{1.88}{\mathrm{Sr}}_{0.12}{\mathrm{CuO}}_{4}$}},\ }\href
  {https://doi.org/10.1103/PhysRevB.90.100510} {\bibfield  {journal} {\bibinfo
  {journal} {Phys. Rev. B}\ }\textbf {\bibinfo {volume} {90}},\ \bibinfo
  {pages} {100510} (\bibinfo {year} {2014})}\BibitemShut {NoStop}%
\bibitem [{\citenamefont {Wen}\ \emph {et~al.}(2019)\citenamefont {Wen},
  \citenamefont {Huang}, \citenamefont {Lee}, \citenamefont {Jang},
  \citenamefont {Knight}, \citenamefont {Lee}, \citenamefont {Fujita},
  \citenamefont {Suzuki}, \citenamefont {Asano}, \citenamefont {Kivelson},
  \citenamefont {Kao},\ and\ \citenamefont {Lee}}]{WenNatComm2019}%
  \BibitemOpen
  \bibfield  {author} {\bibinfo {author} {\bibfnamefont {J.}~\bibnamefont
  {Wen}}, \bibinfo {author} {\bibfnamefont {H.}~\bibnamefont {Huang}}, \bibinfo
  {author} {\bibfnamefont {S.~J.}\ \bibnamefont {Lee}}, \bibinfo {author}
  {\bibfnamefont {H.}~\bibnamefont {Jang}}, \bibinfo {author} {\bibfnamefont
  {J.}~\bibnamefont {Knight}}, \bibinfo {author} {\bibfnamefont {Y.~S.}\
  \bibnamefont {Lee}}, \bibinfo {author} {\bibfnamefont {M.}~\bibnamefont
  {Fujita}}, \bibinfo {author} {\bibfnamefont {K.~M.}\ \bibnamefont {Suzuki}},
  \bibinfo {author} {\bibfnamefont {S.}~\bibnamefont {Asano}}, \bibinfo
  {author} {\bibfnamefont {S.~A.}\ \bibnamefont {Kivelson}}, \bibinfo {author}
  {\bibfnamefont {C.~C.}\ \bibnamefont {Kao}},\ and\ \bibinfo {author}
  {\bibfnamefont {J.-S.}\ \bibnamefont {Lee}},\ }\bibfield  {title} {\bibinfo
  {title} {Observation of two types of charge-density wave orders in
  superconducting $\mathrm{La_{2-x}Sr_{x}CuO_{4}}$},\ }\href@noop {} {\bibfield
   {journal} {\bibinfo  {journal} {Nature Communications}\ }\textbf {\bibinfo
  {volume} {10}},\ \bibinfo {pages} {3269} (\bibinfo {year}
  {2019})}\BibitemShut {NoStop}%
\bibitem [{\citenamefont {Arsenault}\ \emph {et~al.}(2018)\citenamefont
  {Arsenault}, \citenamefont {Takahashi}, \citenamefont {Imai}, \citenamefont
  {He}, \citenamefont {Lee},\ and\ \citenamefont {Fujita}}]{ArsenaultPRB2018}%
  \BibitemOpen
  \bibfield  {author} {\bibinfo {author} {\bibfnamefont {A.}~\bibnamefont
  {Arsenault}}, \bibinfo {author} {\bibfnamefont {S.~K.}\ \bibnamefont
  {Takahashi}}, \bibinfo {author} {\bibfnamefont {T.}~\bibnamefont {Imai}},
  \bibinfo {author} {\bibfnamefont {W.}~\bibnamefont {He}}, \bibinfo {author}
  {\bibfnamefont {Y.~S.}\ \bibnamefont {Lee}},\ and\ \bibinfo {author}
  {\bibfnamefont {M.}~\bibnamefont {Fujita}},\ }\bibfield  {title} {\bibinfo
  {title} {{$^{139}\mathrm{La}$ NMR investigation of the charge and spin order
  in a ${\mathrm{La}}_{1.885}{\mathrm{Sr}}_{0.115}{\mathrm{CuO}}_{4}$ single
  crystal}},\ }\href {https://doi.org/10.1103/PhysRevB.97.064511} {\bibfield
  {journal} {\bibinfo  {journal} {Phys. Rev. B}\ }\textbf {\bibinfo {volume}
  {97}},\ \bibinfo {pages} {064511} (\bibinfo {year} {2018})}\BibitemShut
  {NoStop}%
\bibitem [{\citenamefont {Imai}\ \emph {et~al.}(2017)\citenamefont {Imai},
  \citenamefont {Takahashi}, \citenamefont {Arsenault}, \citenamefont {Acton},
  \citenamefont {Lee}, \citenamefont {He}, \citenamefont {Lee},\ and\
  \citenamefont {Fujita}}]{ImaiPRB2017}%
  \BibitemOpen
  \bibfield  {author} {\bibinfo {author} {\bibfnamefont {T.}~\bibnamefont
  {Imai}}, \bibinfo {author} {\bibfnamefont {S.~K.}\ \bibnamefont {Takahashi}},
  \bibinfo {author} {\bibfnamefont {A.}~\bibnamefont {Arsenault}}, \bibinfo
  {author} {\bibfnamefont {A.~W.}\ \bibnamefont {Acton}}, \bibinfo {author}
  {\bibfnamefont {D.}~\bibnamefont {Lee}}, \bibinfo {author} {\bibfnamefont
  {W.}~\bibnamefont {He}}, \bibinfo {author} {\bibfnamefont {Y.~S.}\
  \bibnamefont {Lee}},\ and\ \bibinfo {author} {\bibfnamefont {M.}~\bibnamefont
  {Fujita}},\ }\bibfield  {title} {\bibinfo {title} {{Revisiting
  $^{63}\mathrm{Cu}$ NMR evidence for charge order in superconducting
  ${\mathrm{La}}_{1.885}{\mathrm{Sr}}_{0.115}{\mathrm{CuO}}_{4}$}},\ }\href
  {https://doi.org/10.1103/PhysRevB.96.224508} {\bibfield  {journal} {\bibinfo
  {journal} {Phys. Rev. B}\ }\textbf {\bibinfo {volume} {96}},\ \bibinfo
  {pages} {224508} (\bibinfo {year} {2017})}\BibitemShut {NoStop}%
\bibitem [{\citenamefont {Curro}\ \emph {et~al.}(2000)\citenamefont {Curro},
  \citenamefont {Hammel}, \citenamefont {Suh}, \citenamefont {H\"ucker},
  \citenamefont {B\"uchner}, \citenamefont {Ammerahl},\ and\ \citenamefont
  {Revcolevschi}}]{Curro}%
  \BibitemOpen
  \bibfield  {author} {\bibinfo {author} {\bibfnamefont {N.~J.}\ \bibnamefont
  {Curro}}, \bibinfo {author} {\bibfnamefont {P.~C.}\ \bibnamefont {Hammel}},
  \bibinfo {author} {\bibfnamefont {B.~J.}\ \bibnamefont {Suh}}, \bibinfo
  {author} {\bibfnamefont {M.}~\bibnamefont {H\"ucker}}, \bibinfo {author}
  {\bibfnamefont {B.}~\bibnamefont {B\"uchner}}, \bibinfo {author}
  {\bibfnamefont {U.}~\bibnamefont {Ammerahl}},\ and\ \bibinfo {author}
  {\bibfnamefont {A.}~\bibnamefont {Revcolevschi}},\ }\bibfield  {title}
  {\bibinfo {title} {Inhomogeneous low frequency spin dynamics in
  ${\mathrm{la}}_{1.65}{\mathrm{eu}}_{0.2}{\mathrm{sr}}_{0.15}{\mathrm{cuo}}_{4}$},\
  }\href {https://doi.org/10.1103/PhysRevLett.85.642} {\bibfield  {journal}
  {\bibinfo  {journal} {Phys. Rev. Lett.}\ }\textbf {\bibinfo {volume} {85}},\
  \bibinfo {pages} {642} (\bibinfo {year} {2000})}\BibitemShut {NoStop}%
\bibitem [{\citenamefont {Mitrovi\'{c}}\ \emph {et~al.}(2008)\citenamefont
  {Mitrovi\'{c}}, \citenamefont {Julien}, \citenamefont {de~Vaulx},
  \citenamefont {Horvati\'{c}}, \citenamefont {Berthier}, \citenamefont
  {Suzuki},\ and\ \citenamefont {Yamada}}]{Mitrovic}%
  \BibitemOpen
  \bibfield  {author} {\bibinfo {author} {\bibfnamefont {V.~F.}\ \bibnamefont
  {Mitrovi\'{c}}}, \bibinfo {author} {\bibfnamefont {M.-H.}\ \bibnamefont
  {Julien}}, \bibinfo {author} {\bibfnamefont {C.}~\bibnamefont {de~Vaulx}},
  \bibinfo {author} {\bibfnamefont {M.}~\bibnamefont {Horvati\'{c}}}, \bibinfo
  {author} {\bibfnamefont {C.}~\bibnamefont {Berthier}}, \bibinfo {author}
  {\bibfnamefont {T.}~\bibnamefont {Suzuki}},\ and\ \bibinfo {author}
  {\bibfnamefont {K.}~\bibnamefont {Yamada}},\ }\bibfield  {title} {\bibinfo
  {title} {Similar glassy features in the $^{139}\text{L}\text{a}$ {NMR}
  response of pure and disordered
  {${\text{La}}_{1.88}{\text{Sr}}_{0.12}{\text{CuO}}_{4}$}},\ }\href
  {https://doi.org/10.1103/PhysRevB.78.014504} {\bibfield  {journal} {\bibinfo
  {journal} {Phys. Rev. B}\ }\textbf {\bibinfo {volume} {78}},\ \bibinfo
  {pages} {014504} (\bibinfo {year} {2008})}\BibitemShut {NoStop}%
\bibitem [{\citenamefont {Gezo}\ \emph {et~al.}(2013)\citenamefont {Gezo},
  \citenamefont {Lui}, \citenamefont {Wolin}, \citenamefont {Slichter},
  \citenamefont {Giannetta},\ and\ \citenamefont {Schlueter}}]{Gezo}%
  \BibitemOpen
  \bibfield  {author} {\bibinfo {author} {\bibfnamefont {J.}~\bibnamefont
  {Gezo}}, \bibinfo {author} {\bibfnamefont {T.-K.}\ \bibnamefont {Lui}},
  \bibinfo {author} {\bibfnamefont {B.}~\bibnamefont {Wolin}}, \bibinfo
  {author} {\bibfnamefont {C.~P.}\ \bibnamefont {Slichter}}, \bibinfo {author}
  {\bibfnamefont {R.}~\bibnamefont {Giannetta}},\ and\ \bibinfo {author}
  {\bibfnamefont {J.~A.}\ \bibnamefont {Schlueter}},\ }\bibfield  {title}
  {\bibinfo {title} {Stretched exponential spin relaxation in organic
  superconductors},\ }\href {https://doi.org/10.1103/PhysRevB.88.140504}
  {\bibfield  {journal} {\bibinfo  {journal} {Phys. Rev. B}\ }\textbf {\bibinfo
  {volume} {88}},\ \bibinfo {pages} {140504} (\bibinfo {year}
  {2013})}\BibitemShut {NoStop}%
\bibitem [{\citenamefont {Dioguardi}\ \emph {et~al.}(2015)\citenamefont
  {Dioguardi}, \citenamefont {Lawson}, \citenamefont {Bush}, \citenamefont
  {Crocker}, \citenamefont {Shirer}, \citenamefont {Nisson}, \citenamefont
  {Kissikov}, \citenamefont {Ran}, \citenamefont {Bud'ko}, \citenamefont
  {Canfield}, \citenamefont {Yuan}, \citenamefont {Kuhns}, \citenamefont
  {Reyes}, \citenamefont {Grafe},\ and\ \citenamefont {Curro}}]{Dioguardi}%
  \BibitemOpen
  \bibfield  {author} {\bibinfo {author} {\bibfnamefont {A.~P.}\ \bibnamefont
  {Dioguardi}}, \bibinfo {author} {\bibfnamefont {M.~M.}\ \bibnamefont
  {Lawson}}, \bibinfo {author} {\bibfnamefont {B.~T.}\ \bibnamefont {Bush}},
  \bibinfo {author} {\bibfnamefont {J.}~\bibnamefont {Crocker}}, \bibinfo
  {author} {\bibfnamefont {K.~R.}\ \bibnamefont {Shirer}}, \bibinfo {author}
  {\bibfnamefont {D.~M.}\ \bibnamefont {Nisson}}, \bibinfo {author}
  {\bibfnamefont {T.}~\bibnamefont {Kissikov}}, \bibinfo {author}
  {\bibfnamefont {S.}~\bibnamefont {Ran}}, \bibinfo {author} {\bibfnamefont
  {S.~L.}\ \bibnamefont {Bud'ko}}, \bibinfo {author} {\bibfnamefont {P.~C.}\
  \bibnamefont {Canfield}}, \bibinfo {author} {\bibfnamefont {S.}~\bibnamefont
  {Yuan}}, \bibinfo {author} {\bibfnamefont {P.~L.}\ \bibnamefont {Kuhns}},
  \bibinfo {author} {\bibfnamefont {A.~P.}\ \bibnamefont {Reyes}}, \bibinfo
  {author} {\bibfnamefont {H.-J.}\ \bibnamefont {Grafe}},\ and\ \bibinfo
  {author} {\bibfnamefont {N.~J.}\ \bibnamefont {Curro}},\ }\bibfield  {title}
  {\bibinfo {title} {{NMR} evidence for inhomogeneous glassy behavior driven by
  nematic fluctuations in iron arsenide superconductors},\ }\href
  {https://doi.org/10.1103/PhysRevB.92.165116} {\bibfield  {journal} {\bibinfo
  {journal} {Phys. Rev. B}\ }\textbf {\bibinfo {volume} {92}},\ \bibinfo
  {pages} {165116} (\bibinfo {year} {2015})}\BibitemShut {NoStop}%
\bibitem [{\citenamefont {Butler}\ \emph {et~al.}(1981)\citenamefont {Butler},
  \citenamefont {Reeds},\ and\ \citenamefont {Dawson}}]{butler:SIAM1981}%
  \BibitemOpen
  \bibfield  {author} {\bibinfo {author} {\bibfnamefont {J.~P.}\ \bibnamefont
  {Butler}}, \bibinfo {author} {\bibfnamefont {J.~A.}\ \bibnamefont {Reeds}},\
  and\ \bibinfo {author} {\bibfnamefont {S.~V.}\ \bibnamefont {Dawson}},\
  }\bibfield  {title} {\bibinfo {title} {Estimating solutions of first kind
  integral equations with non-negative constraints and optimal smoothing},\
  }\href@noop {} {\bibfield  {journal} {\bibinfo  {journal} {SIAM J. Numer.
  Anal.}\ }\textbf {\bibinfo {volume} {18(3)}},\ \bibinfo {pages} {381}
  (\bibinfo {year} {1981})}\BibitemShut {NoStop}%
\bibitem [{\citenamefont {Lawson}\ and\ \citenamefont
  {Hanson}(1974)}]{lawson:1974}%
  \BibitemOpen
  \bibfield  {author} {\bibinfo {author} {\bibfnamefont {C.}~\bibnamefont
  {Lawson}}\ and\ \bibinfo {author} {\bibfnamefont {R.}~\bibnamefont
  {Hanson}},\ }\href@noop {} {\emph {\bibinfo {title} {Solving Least Squares
  Problems}}}\ (\bibinfo  {publisher} {Prentice-Hall, Englewood Cliffs, NJ},\
  \bibinfo {year} {1974})\BibitemShut {NoStop}%
\end{thebibliography}

%


\end{document}